\documentclass[apj]{emulateapj}
\usepackage{apjfonts}

\newcommand{\be}{\begin{eqnarray}} 
\newcommand{\ee}{\end{eqnarray}}

\begin{document}
 
\title{Long Type I X-ray Bursts and Neutron Star Interior Physics}
 
\author{Andrew Cumming\altaffilmark{1}, Jared Macbeth\altaffilmark{2}, J.~J.~M.~in 't Zand\altaffilmark{3}, and Dany Page\altaffilmark{4}}
\altaffiltext{1}{Physics Department, McGill University, 3600 rue University, Montreal, QC, H3A 2T8, Canada}
\altaffiltext{2}{Department of Astronomy and Astrophysics, University of California, Santa Cruz, CA 95064, USA}
\altaffiltext{3}{SRON National Institute for Space Research, Sorbonnelaan 2, NL-3584 CA Utrecht, The Netherlands}
\altaffiltext{4}{Instituto de Astronom\'ia, Universidad Nacional Aut\'onoma de M\'exico, 04510 Mexico D.F., Mexico}

\begin{abstract}
Superbursts are very energetic Type I X-ray bursts discovered in recent years by long term monitoring of X-ray bursters, believed to be due to unstable ignition of carbon in the deep ocean of the neutron star. A number of intermediate duration bursts have also been observed, probably associated with ignition of a thick helium layer. We investigate the sensitivity of these long X-ray bursts to the thermal profile of the neutron star crust and core. We first compare cooling models of superburst lightcurves with observations, and derive constraints on the ignition mass and energy release. Despite the large uncertainties associated with the distance to each source, these parameters are quite well constrained in our fits. For the observed superbursts, we find ignition column depths in the range $0.5$--$3\times 10^{12}\ {\rm g\ cm^{-2}}$, and energy release $\approx 2\times 10^{17}\ {\rm erg\ g^{-1}}$. This energy release implies carbon fractions of $X_C>10$\%, constraining models of rp-process hydrogen burning. We then calculate ignition models for superbursts and pure helium bursts, and compare to observations. We show that achieving unstable ignition of carbon at accretion rates less than 0.3 of the Eddington rate requires $X_C\gtrsim 0.2$, consistent with our lightcurve fits. Most importantly, we find that when Cooper pairing neutrino emission in the crust is included, the crust temperature is too low to support unstable carbon ignition at column depths of $\sim 10^{12}\ {\rm g\ cm^{-2}}$. Some additional heating mechanism is required in the accumulating fuel layer to explain the observed properties of superbursts. If Cooper pair emission is less efficient than currently thought, the observed ignition depths for superbursts imply that the crust is a poor conductor, and  the core neutrino emission is not more efficient than modified URCA. The observed properties of helium bursts support these conclusions, requiring inefficient crust conductivity and core neutrino emission.
\end{abstract}


\section{Introduction}

In the past few years a new regime of nuclear burning on the surfaces of accreting neutron stars has been revealed by the discovery of X-ray superbursts (Cornelisse et al.~2000; Strohmayer \& Brown 2002; Kuulkers 2003). These are rare (recurrence times $\approx 1$ year), extremely energetic (energies $\approx 10^{42}\ {\rm ergs}$) and long duration ($4$--$14$ hours) Type I X-ray bursts that have been discovered with long term monitoring campaigns by BeppoSAX and RXTE. Whereas normal Type I X-ray bursts involve unstable thermonuclear ignition of hydrogen and helium (see Lewin, van Paradijs, \& Taam 1993, 1995; Strohmayer \& Bildsten 2003 for reviews), superbursts are thought to involve ignition of carbon in a much thicker layer (Cumming \& Bildsten 2001, hereafter CB01; Strohmayer \& Brown 2002). 

Theoretical studies of superbursts initially focused on their potential as probes of nuclear physics. The fuel for superbursts is thought to be produced by hydrogen and helium burning at lower densities by the rp-process (Wallace \& Woosley 1981), a series of proton captures and beta-decays on heavy nuclei close to the proton drip line. This process naturally explains the $\approx 100\ {\rm s}$ extended tails observed in some X-ray bursts (e.g.~from the regular burster GS~1826-24; Galloway et al.~2004). The amount of carbon remaining after H/He burning depends on the details of the rp-process (Schatz et al.~2003b), which involves unstable heavy nuclei whose properties are not well known experimentally (Schatz et al.~1998).

CB01 argued that the heavy elements made by the rp-process will make the accumulating layer less conductive to heat, increasing the temperature gradient within it, and leading to earlier ignition than pure carbon models (as had been considered earlier by Woosley \& Taam 1976, Taam \& Picklum 1978, Lamb \& Lamb 1978, and Brown \& Bildsten 1998), in better agreement with observed superburst energies. In fact, Brown (2004) and Cooper \& Narayan (2005) showed that the ignition conditions are  much more sensitive to the thermal properties of the neutron star interior, specifically the neutrino emissivity of the neutron star core and composition of the crust. This is exciting because it offers a new way to probe the neutron star interior, complementary to observations of transiently-accreting neutron stars in quiescence (Brown et al.~1998, Colpi et al.~2001, Rutledge et al.~2002; Wijnands et al.~2002; Yakovlev et al.~2004), or cooling isolated neutron stars (see Yakovlev \& Pethick 2004 and Page et al.~2005 for recent reviews).

Several other long duration X-ray bursts have been observed that are intermediate in duration and energy between normal Type I X-ray bursts and superbursts (e.g.~Figure 1 of Kuulkers 2003). These intermediate bursts have durations of $\approx 30\ {\rm mins}$ and energies $\approx 10^{41}\ {\rm ergs}$, and sources include SLX~1737-282 (in 't Zand et al.~2002), 1RXS~J171824.2-402934 (Kaptein et al.~2000), and 2S~0918-549 (in 't Zand et al.~2005). Long duration bursts are expected from accretion of hydrogen and helium at low rates $\approx 0.01\ \dot M_{\rm Edd}$ (Fujimoto, Hanawa, \& Miyaji 1981; Bildsten 1998; Narayan \& Heyl 2003; Cumming 2003b). For accretion of solar composition material at these accretion rates, a massive layer of pure helium accumulates and ignites beneath a steady hydrogen burning shell. Whereas the hydrogen shell is heated by hot CNO hydrogen burning, the helium shell is heated mainly by the heat flux emerging from the crust. Therefore, just like superbursts, these bursts are potentially sensitive to the crust composition and core temperature. The case of pure helium accretion is particularly interesting because heating by hydrogen burning then plays no role, making the ignition conditions directly sensitive to interior physics.

It is therefore important to constrain the ignition depth, recurrence times, and energy released during superbursts and other long X-ray bursts. Our knowledge of superburst recurrence times is limited because they are rare events. Three superbursts were seen from 4U~1636-54 separated by intervals of 2.9 and 1.8 years (Wijnands 2001; Kuulkers et al.~2004). Dividing the total duration of observations of X-ray bursters with the BeppoSAX/WFC by the number of superbursts observed gives a recurrence time estimate of $0.4$--$2$ years (in 't Zand et al.~2003). Brown (2004) and Cooper \& Narayan (2005) emphasized that to achieve ignition of carbon on $\approx 1\ {\rm year}$ timescales at accretion rates $\dot M\approx 0.1\ \dot M_{\rm Edd}$ requires the accumulating layer to be sufficiently hot. An enhanced core neutrino emissivity (as would be produced if, for example, the direct URCA process operated in the core, e.g.~Yakovlev \& Pethick 2004, Page et al.~2005) together with a large crust conductivity gives very long ($\gg 10\ {\rm yr}$) superburst recurrence times, inconsistent with observations. Recently, in~'t Zand et al.~(2005) showed that the intermediate duration X-ray burst from 2S~0918-549 is well explained by accretion of pure helium at the observed rate of $\dot M\approx 0.01\ \dot M_{\rm Edd}$, assuming that most of the heat released in the crust by pycnonuclear reactions and electron captures (Haensel \& Zdunik 1990, 2003) flows outwards and heats the accumulating helium layer. However, they did not explore the sensitivity of this assumption to the interior physics.

In this paper, we investigate the constraints on interior physics coming from superbursts and pure helium bursts.  We first derive independent constraints on the ignition depth and energetics of superbursts by fitting the observed lightcurves to theoretical cooling models as calculated by Cumming \& Macbeth (2004, hereafter CM04). We then calculate ignition conditions for both superbursts and pure helium bursts, and compare with observed properties. We start in \S \ref{sec:cool} by summarizing the properties of our cooling models for superbursts, present the fits to the observed lightcurves, and discuss the constraints on the ignition depth and energy release. In \S \ref{sec:ignite}, we calculate ignition conditions for superbursts and discuss the implications for the thermal structure of the interior. We show that the best fit is obtained for inefficient neutrino emission in the neutron star crust and core. Next, in \S \ref{sec:helium}, we apply these ignition models to pure helium bursts, and show that their properties imply the same conclusion: inefficient neutrino emission. We conclude in \S \ref{sec:conc}. In Appendix A, we discuss a simple model of the early phase of the superburst lightcurve which reproduces the time-dependent results, and in Appendix B give  approximate analytic solutions for the crust temperature profile.


\section{Cooling models for superbursts and comparison to observations}
\label{sec:cool}

\subsection{Cooling models}

CM04 computed lightcurves for superbursts by assuming that the fuel is burned locally and instantly at each depth, and then following the thermal evolution and surface luminosity as the burning layers cool. They showed that the lightcurve of the cooling tail of the superburst is a broken power law, with time of the break proportional to the cooling time of the entire layer. The early phase of cooling depends mostly on the energy released in the flash; the late phase of cooling depends mostly on the thickness of the layer. We now apply these models to the observations, and discuss the constraints on superburst ignition conditions.

We refer the reader to CM04 for full details of the cooling models, including approximate analytic expressions for the flux as a function of time. The parameters of the models are the ignition column depth ($y_{12}\ 10^{12}\ {\rm g\ cm^{-2}}$), and the energy release per gram ($E_{17}\ 10^{17}\ {\rm erg\ g^{-1}}$) which is assumed to be independent of depth. The thermal evolution is followed numerically by the method of lines, which involves finite differencing on a spatial grid, and then integrating the resulting set of coupled ordinary differential equations for the temperature at each grid point forward in time. It is important to note that the rise of the superburst is not resolved, since the entire fuel layer is assumed to burn instantly. The models assume a neutron star mass and radius of $M=1.4\ M_\odot$ and $R=10\ {\rm km}$, giving the surface gravity $g=(GM/R^2)(1+z)=2.45\times 10^{14}\ {\rm cm\ s^{-2}}$ and redshift $1+z=1.31$. We include the equation of state, radiative and conductive opacities, heat capacity, and neutrino emissivities as described by Schatz et al.~(2003a). At the temperatures and densities appropriate for a superburst, the neutrino emission is mostly due to pair annhilation (CM04).

 \begin{figure}
\epsscale{1.2}
\plotone{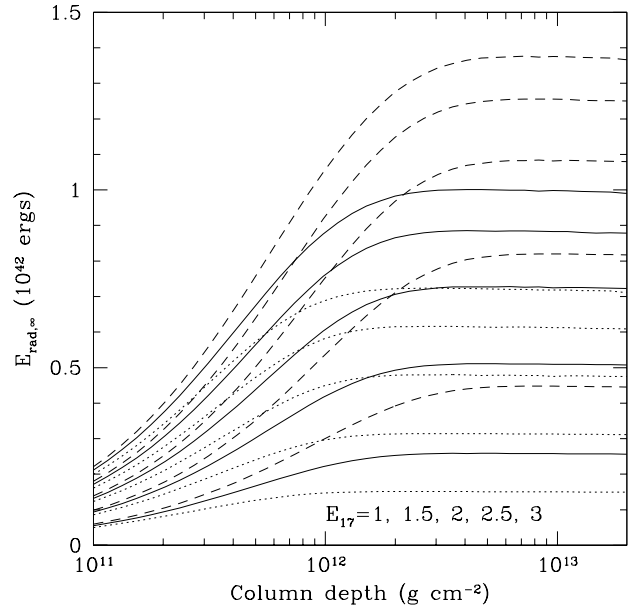}
\caption{Energy radiated from the surface after 3 hours ({\em dotted curves}), 6 hours ({\em solid curves}), and 12 hours ({\em dashed curves}) for $E_{17}=1,1.5,2,2.5$ and $3$, as a function of the column depth. We assume a neutron star radius $R=10$ km ($E_{{\rm rad},\infty}\propto R^2$).\label{fig:er2}}
\end{figure}

\begin{figure}
\epsscale{1.1}
\plotone{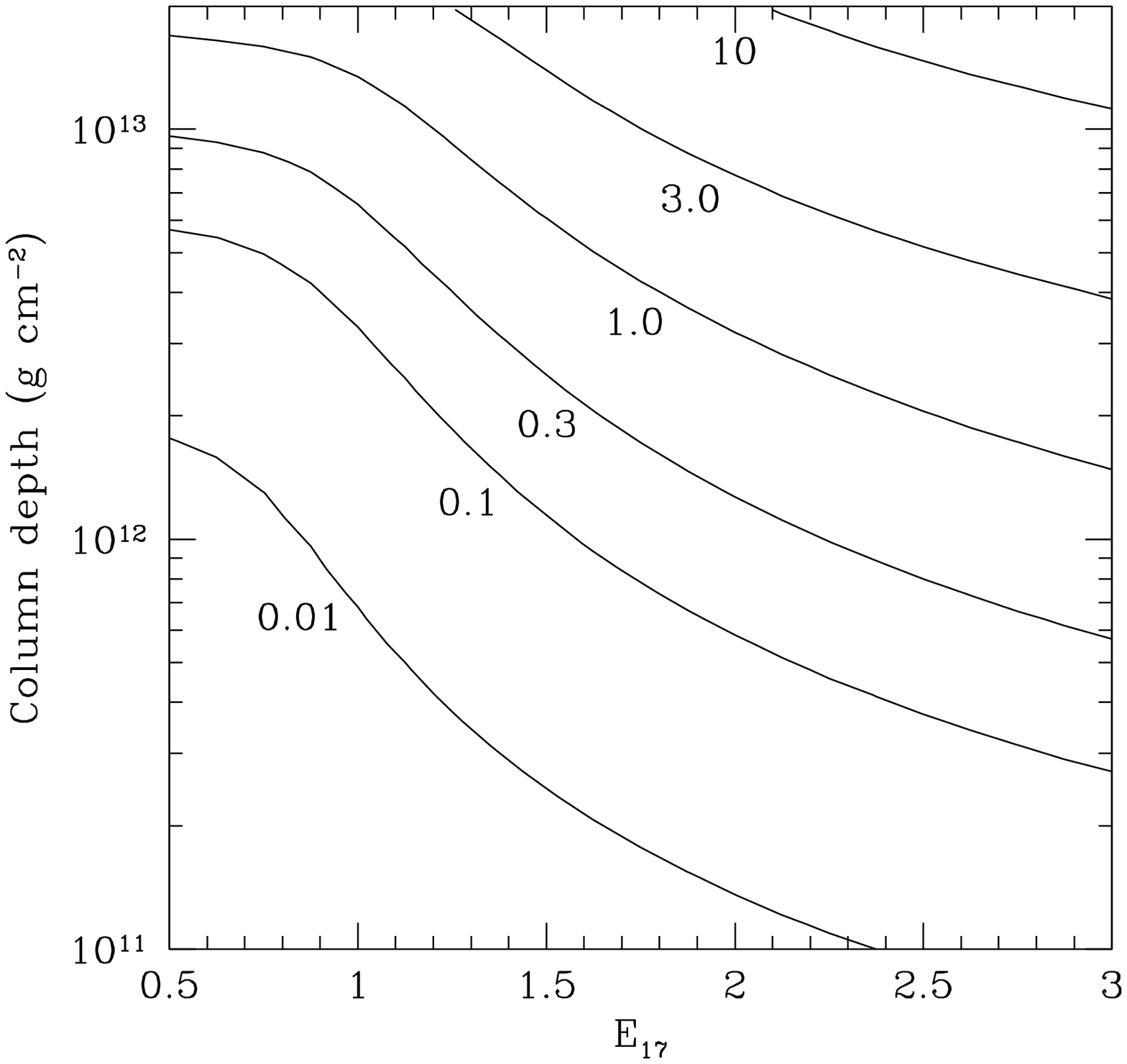}
\plotone{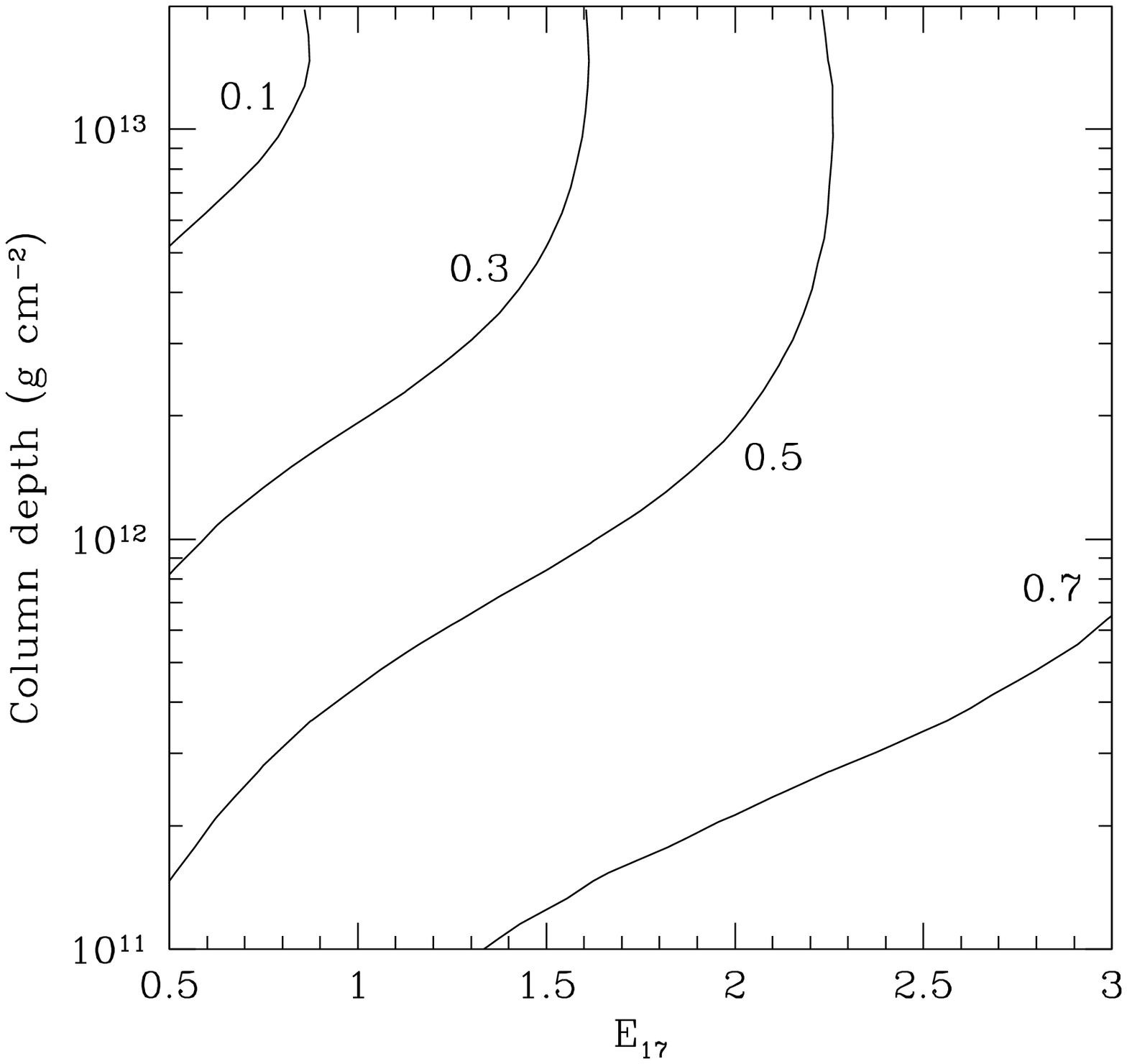}
\caption{The ``thermostats'' of neutrino emission and inwards conduction of heat. {\em Upper panel:} ratio of energy released as neutrinos to energy radiated from the surface, both in the first 24 hours. Neutrinos dominate the energy release for large $y$ and $E_{17}$. {\em Lower panel:} Fraction of the total nuclear energy released that escapes in the first 24 hours, either as neutrinos or from the surface. For large $y$, a significant fraction of the energy released is conducted inwards and released on a longer timescale.\label{fig:loss}}
\end{figure}

We first summarize some of the properties of the models. CM04 already noted that the power law decay gives a long tail to the superburst lightcurves, which is similar to the long tails observed in superbursts (Kuulkers et al.~2002; Cornelisse et al.~2002). Figure \ref{fig:er2} shows the amount of energy radiated from the surface in the first 3, 6, and 12 hours as a function of column depth for different choices of $E_{17}$. The insensitivity of radiated energy to column depth for $y\gtrsim 10^{12}\ {\rm g\ cm^{-2}}$ in Figure \ref{fig:er2} shows that the total emitted energy is not a good indicator of the ignition column depth. There are two reasons for the characteristic radiated energy of $\approx 10^{42}\ {\rm ergs}$ (Strohmayer \& Brown 2002). First, neutrino emission takes away most of the energy for large columns, and secondly heat flows inwards to be released on longer timescales. These effects are quantified as a function of $y_{12}$ and $E_{17}$ in Figure \ref{fig:loss}. The first panel shows the ratio of energy lost as neutrinos to the energy lost through the surface. For example, for $y\approx 10^{13}\ {\rm g\ cm^{-2}}$ and $E_{17}\approx 3$, neutrinos take away an order of magnitude more energy than is lost from the surface. This is in rough agreement with the one-zone model of Strohmayer \& Brown (2002). The second panel shows the fraction of energy that is lost in the first 24 hours, either as neutrinos, or from the surface. The remaining energy, which is released on longer timescales, can be a significant fraction of the total for column depths $\approx 10^{13}\ {\rm g\ cm^{-2}}$ and $E_{17}\approx 1$.

Even without detailed fits to observed lightcurves, these results give some indication of the values of $E_{17}$ needed to match the observed properties of superbursts. Figure \ref{fig:er2} shows that an energy release $E_{17}>1$ is required for the observed burst energy to reach $\gtrsim 3\times 10^{41}\ {\rm ergs}$ during the first few hours. On the other hand, for large values of $E_{17}$, the initial flux exceeds the Eddington flux $F_{\rm Edd}=cg/\kappa=2.2\times 10^{25}\ {\rm erg\ cm^{-2}\ s^{-1}}\ (g_{14}/2.45)(1.7/(1+X))$, in which case the superburst would be expected to show photospheric radius expansion. In Figure \ref{fig:edd}, we plot the time for which the flux exceeds $F_{\rm Edd}$ for different $E_{17}$ values. This time is not very sensitive to the ignition column, since the early evolution of the burst is independent of the layer thickness. For $E_{17}\gtrsim 2$, the flux is super-Eddington for timescales of minutes or longer. The superburst from 4U~1820-30 showed an extended period of photospheric radius expansion lasting for $\sim 1000\ {\rm s}$ (SB02). This is in good agreement with the expectation that this source had a significant energy release due to large amounts of carbon produced by stable burning of pure helium (SB02; Cumming 2003a). Figure \ref{fig:edd} implies that $E_{17}>5$ is required to get such a long period of super-Eddington luminosity with pure helium. However, there is no strong evidence for photospheric radius expansion in any other superburst\footnote{Precursors were seen with BeppoSAX/WFC from KS~1731-260, 4U~1254-69, and GX~17+2. In GX~17+2, the spectral data are of insufficient quality to see radius expansion during the precursor or the minutes thereafter (because of the high persistent emission); in KS~1731-260, no radius expansion was seen (Kuulkers et al.~2002); in 4U 1254-69 there are no indications. In all cases the peak flux of the precursor is smaller by a factor of $1.5$--$2$ than the brightest of the ordinary bursts.}. Taken together, these two constraints imply that  $E_{17}\approx 2$ for most superbursts.

\subsection{Fits to superburst lightcurves}

We have fitted the superburst lightcurves to the cooling models. The parameters of the models are $E_{17}$ and $y_{12}$. However, there are two additional parameters in our fits. The rise of most superbursts is not observed because of data gaps, making the start time of the burst uncertain, and so we include the start time as an extra parameter. Most importantly, the distance to the source is not well constrained in most cases. This dominates the uncertainty in the fitted parameters, and so we have fitted the models by holding distance fixed at different trial values, and searching over the remaining parameters to find the best fitting model at each distance. We include BeppoSAX/WFC data for the superbursts from 4U 1254-690 (in 't Zand et al.~2003), KS 1731-260 (Kuulkers et al.~2002), 4U 1735-444 (Cornelisse et al.~2000), Ser X-1 (Cornelisse et al.~2002), GX~17+2 (in't Zand et al.~2004), and the RXTE/PCA lightcurve of 4U~1636-54 (Strohmayer \& Markwardt 2002; Kuulkers et al.~2004). For GX~17+2, we use burst 2 from Figure 7 of in 't Zand et al.~(2004). This is one of the best candidates for a superburst, and has the most complete lightcurve.

\begin{figure}
\epsscale{1.1}
\plotone{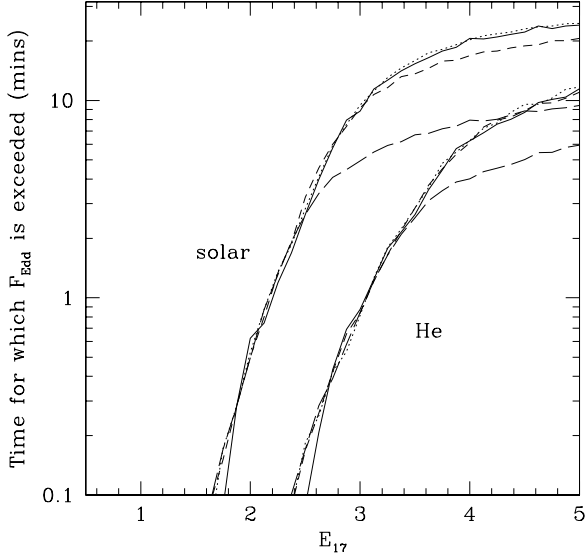}
\caption{Time for which the flux exceeds the Eddington flux as a function of energy release  $E_{17}$. The curves are for $y=10^{11}$ (long-dashed), $3\times 10^{11}$ (short-dashed), $10^{12}$ (dotted), and $10^{13}\ {\rm g\ cm^{-2}}$ (solid). We show two sets of curves for solar composition ($F_{\rm Edd}=2.2\times 10^{25}\ {\rm erg\ cm^{-2}\ s^{-1}}$) and pure helium ($F_{\rm Edd}=3.7\times 10^{25}\ {\rm erg\ cm^{-2}\ s^{-1}}$).\label{fig:edd}}
\end{figure}

\begin{figure*}
\epsscale{0.8}
\plottwo{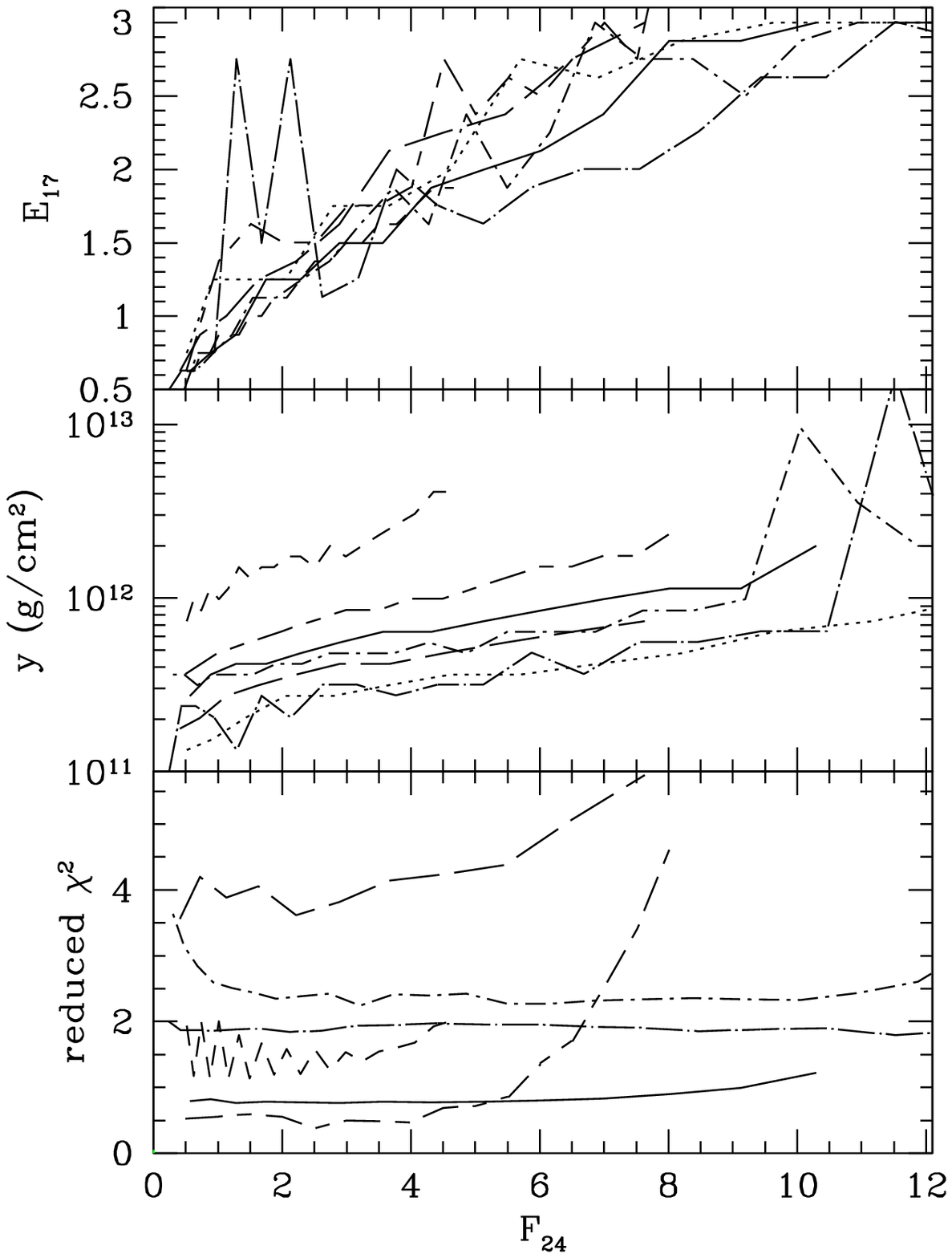}{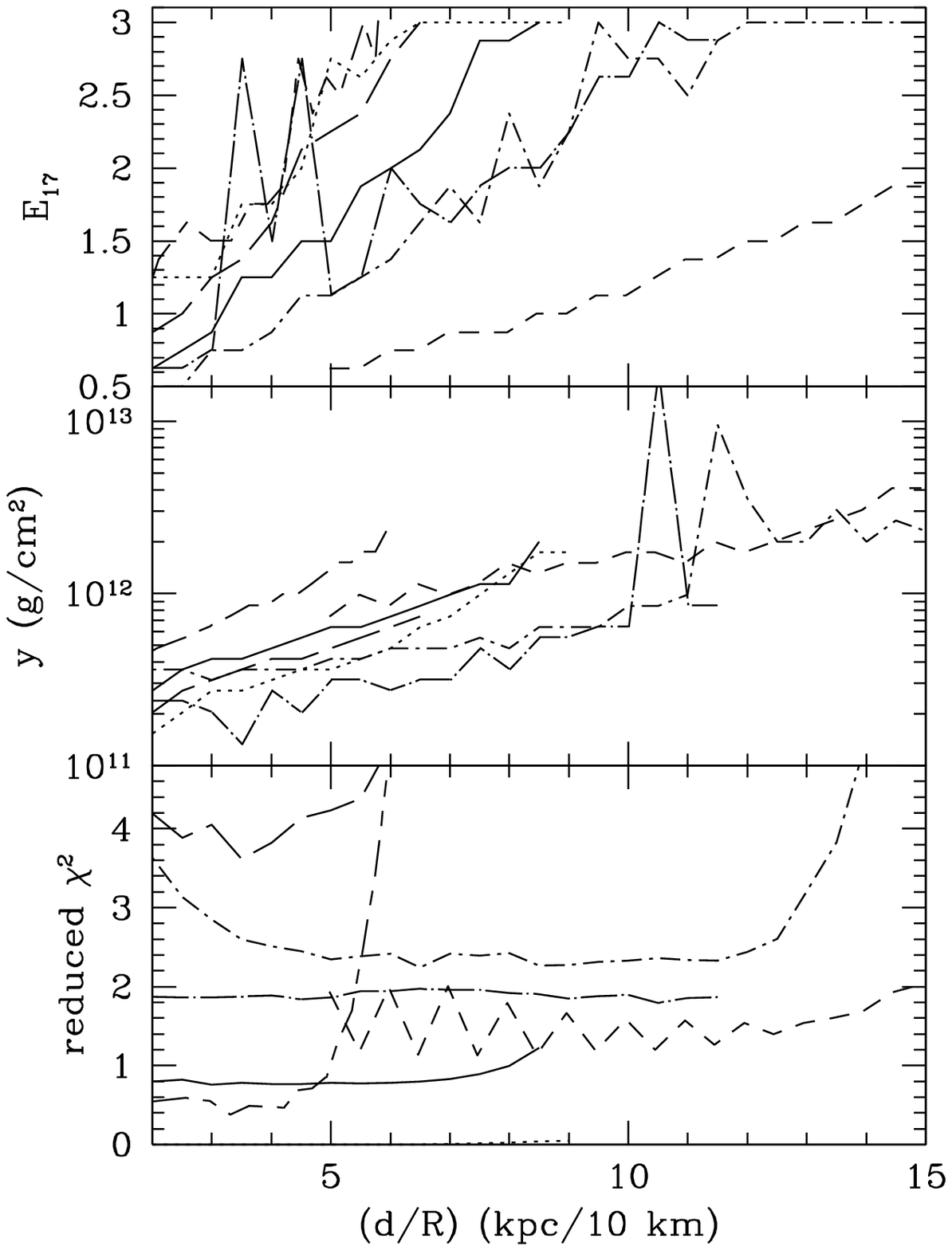}
\caption{{\em Left panel:} best fitting $E_{17}$ and $y$, and the associated reduced $\chi^2$, as a function of assumed peak flux $F_{24}$. The fitted values approximately follow the scalings $E_{17}\approx 0.8 F_{24}^{4/7}$ and $y\propto F_{24}^{5/7}$. We show results for 4U~1254-690 (short-dashed), KS~1731-260 (long dashed-short dashed), 4U~1735-444 (solid), Ser X-1 (long-dashed), GX~17+2 (burst 2 dot-dashed, burst 3 long-dot-dashed), and 4U~1636-54 (dotted). {\em Right panel:} same as left panel, but now using the observed peak flux to plot everything in terms of the distance to the source. The $\chi^2$ for 4U~1636-54 (dotted curves) is off scale in the lower panel.
\label{fig:fit2}}
\end{figure*}

We have extended the CM04 models to a large grid in $E_{17}$ and $y_{12}$ for comparison to the observations. For a given source distance, we calculate the flux at the surface of the star $F_\star$ which corresponds to the observed peak flux $f_{\rm peak}$, i.e.~$4\pi R^2F_{\star}=4\pi d^2f_{\rm peak}$. We will refer to $F_\star$ in units of $10^{24}\ {\rm erg\ cm^{-2}\ s^{-1}}$ as $F_{24}$ . This quantity sets the normalization scale for comparison with the theoretical models, and is  given by
\begin{equation}
F_{24}=9.5\ \left({f_{\rm peak}\over 10^{-8}\ {\rm erg\ cm^{-2}\ s^{-1}}}\right)\left({d/R\over 10\ {\rm kpc}/10\ {\rm km}}\right)^2.
\end{equation}
We then search for the minimum value of $\chi^2$ over the grid of theoretical models, with $E_{17}$ ranging from $0.5$ to $3$ in steps of $0.125$, and $y_{12}$ ranging from $10^{11}$ to $3\times 10^{13}\ {\rm g\ cm^{-2}}$ in steps of $1/16$ in $\log_{10} y$ (i.e.~factors of 15\% between successive $y$ values). For each model, we vary the start time of the superburst between the limits allowed by the observations to find the best fit. Before comparing the model to the data, we redshift the time and flux assuming a gravitational redshift factor of $1+z=1.31$, appropriate for a 10 km, $1.4\ M_\odot$ neutron star. Our results are not very sensitive to variations in the redshift factor within the expected range (roughly $1.2$--$1.5$).

\begin{deluxetable}{lllll}
\tablewidth{0pt}
\tablecaption{Fits to superburst lightcurves\label{tab:fits}}
\tablehead{\colhead{Source} & \colhead{$f_{\rm peak}$\tablenotemark{a}} & \colhead{$d/R$\tablenotemark{b}} & \colhead{$E_{17}$\tablenotemark{c}}
& \colhead{$y_{12}$\tablenotemark{c}}\nl
\colhead{} & \colhead{} & \colhead{} & \colhead{} & \colhead{}
}
\startdata
4U~1254-690 & 0.22 & 13 & 1.5 & 2.7\nl
4U~1735-444 & 1.5 & 8 & 2.6 & 1.3\nl
KS~1731-260 & 2.4	& 4.5 & 1.9 & 1.0\nl
GX~17+2 burst 2 & 0.8 & 8 & 1.8 & 0.64 \nl
Ser~X-1 & 1.9 & 6 & 2.3 & 0.55\nl
4U~1636-54 & 2.4 & 5.9 & 2.6 & 0.48
\enddata
\tablenotetext{a}{Observed peak flux in units of $10^{-8}\ {\rm erg\ cm^{-2}\ s^{-1}}$.}
\tablenotetext{b}{Adopted distance in units of ${\rm kpc}/10\ {\rm km}$.}
\tablenotetext{c}{The fitted parameters scale roughly as $E_{17}\propto (d/R)^{8/7}$ and $y_{12}\propto (d/R)^{10/7}$ (see text). For a 50\% distance uncertainty, the uncertainties in $E_{17}$ and $y_{12}$ are 60\% and 70\% respectively (see also Fig.~\ref{fig:fit2}).}
\end{deluxetable}

Figure \ref{fig:fit2} shows the best fitting $E_{17}$, $y$, and the reduced $\chi^2$ of the fit for each source, as a function of both $F_{24}$ and the distance to radius ratio $d/R$. A larger flux normalization for the observed lightcurve results in larger values of $E_{17}$ and $y_{12}$, which increase in such a way as to maintain the overall shape of the cooling curve. The scalings are straightforward to understand from the analytic expressions for the flux given by CM04 (see eq.~[4] of that paper). At early times, the flux is $F\propto t^{-0.2}E_{17}^{7/4}$ (independent of column depth). In Appendix A, we discuss the physics underlying these scalings. Comparing with the fitted values to set the constant, we find $E_{17}\approx 0.8 F_{24}^{4/7}$, which is in good agreement with the observed relation between the fitted $E_{17}$ value and $F_{24}$. There is a similar scaling for the best fit column depth, which can also be understood from the analytic fit, but now at late times, where $F\propto y E_{17}^{1/2}$, giving $y \propto F^{5/7}$. For a given fit, the fractional uncertainties in $E_{17}$ and $y$ can therefore be estimated as $ \approx (4/7)(\delta F/F)$ and $(5/7)(\delta F/F)$ assuming that the distance uncertainty dominates.

\begin{figure}
\epsscale{1.1}
\plotone{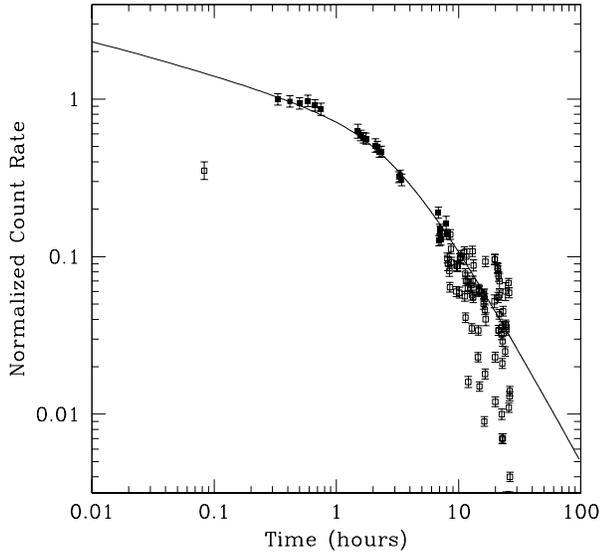}
\caption{Fitted lightcurve for KS~1731-260, assuming the distance given in Table 1. Solid data points are included in the fit, open data points (with fluxes less than 0.1 of the peak flux) are not included. 
\label{fig:lc}}
\end{figure}

\begin{figure}
\epsscale{1.1}
\plotone{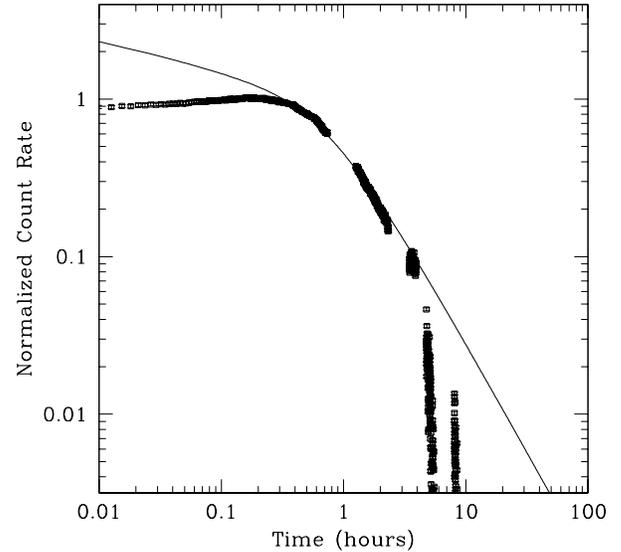}
\caption{Fitted lightcurve for 4U~1636-54.
\label{fig:lc2}}
\end{figure}

\begin{figure}
\epsscale{1.1}
\plotone{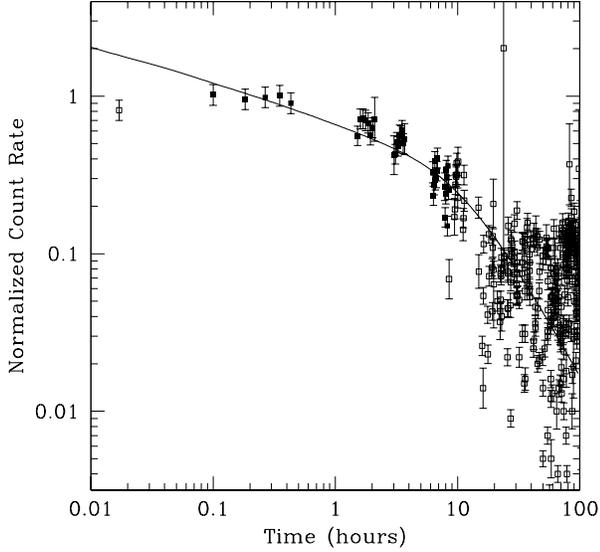}
\caption{Fitted lightcurve for 4U~1254-690.
\label{fig:lc3}}
\end{figure}

\begin{figure}
\epsscale{1.1}
\plotone{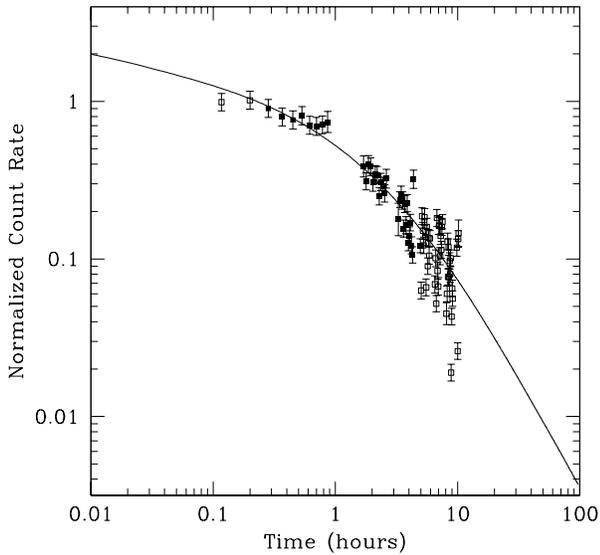}
\caption{Fitted lightcurve for 4U~1735-444.
\label{fig:lc4}}
\end{figure}

\begin{figure}
\epsscale{1.1}
\plotone{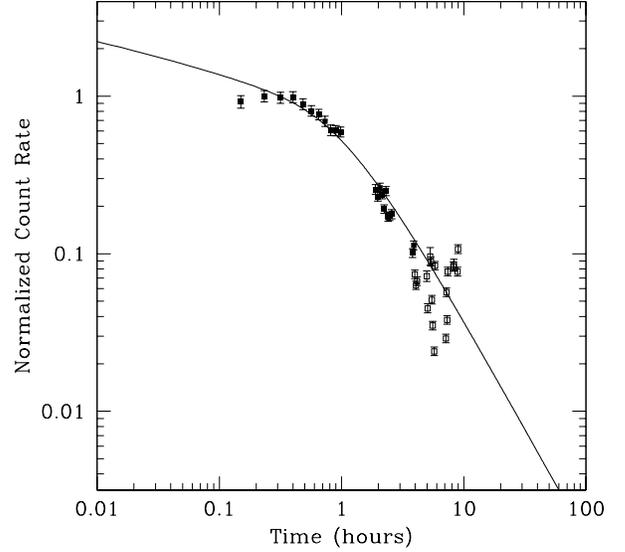}
\caption{Fitted lightcurve for Ser~X-1.
\label{fig:lc5}}
\end{figure}

\begin{figure}
\epsscale{1.1}
\plotone{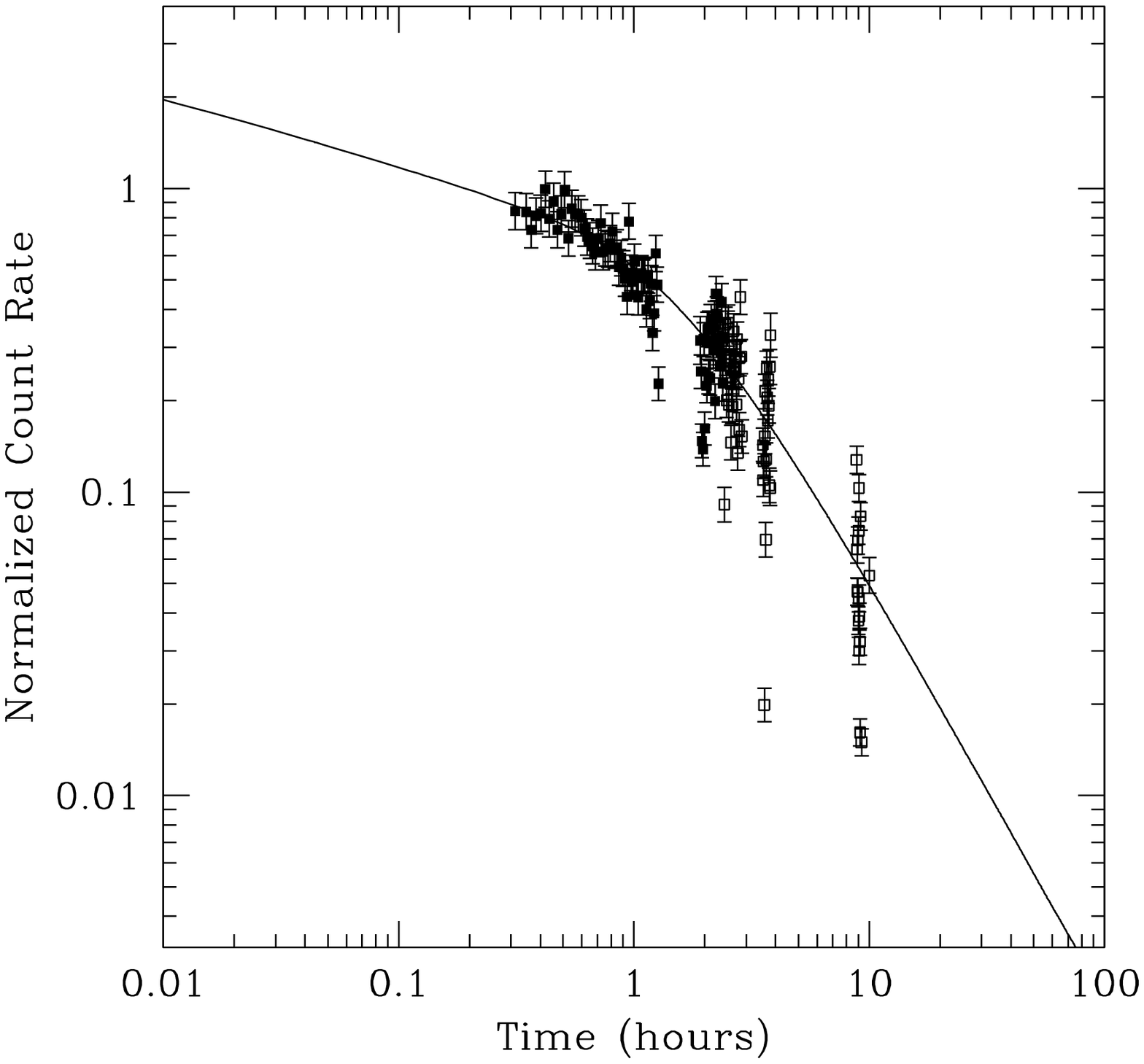}
\caption{Fitted lightcurve for GX~17+2 (burst 2 from in 't Zand et al.~2004).
\label{fig:lc6}}
\end{figure}

For specific choices of distance to each source, we show the best fitting models in Figures 5 to 10, and list the parameters in Table \ref{tab:fits}.  The fitted values can be rescaled to a different distance using the analytic scalings, or by referring to Figure 4. We adopt distance estimates from the literature for 4U~1254-690 (in 't Zand et al.~2003) and 4U~1636-54 (Augustein et al.~1998). For GX~17+2 and 4U~1735-444 we adopt a fiducial value of 8 kpc. For Ser~X-1 and KS~1731-260, we take a lower distance then the upper limits or estimates in the literature, because this significantly improves the fit of our models. For example, Muno et al.~(2000) place a distance limit of $d<7\ {\rm kpc}$ for KS~1731-260 using radius expansion X-ray bursts, assuming that the peak luminosity is the Eddington luminosity for pure helium. We find that for $d/R\gtrsim 5 \ {\rm kpc/10\ km}$ the superburst lightcurve is not well fit by our models, with $\chi^2$ rapidly increasing for larger $d/R$. If the distance is $7\ {\rm kpc}$, the required neutron star radius is $\gtrsim 13\ {\rm km}$. Alternatively, the source could be closer. For example, using the Eddington luminosity for a solar composition rather than pure helium gives a closer distance by a factor of $(1.7)^{1/2}$ or $1.3$. We choose a distance $d/R=4.5\ {\rm kpc/10\ km}$ for the fit shown in Figure \ref{fig:lc}. 

The most detailed lightcurve is for 4U~1636-54. This source has shown three superbursts (Wijnands 2001; Strohmayer \& Markwardt 2002; Kuulkers et al.~2004), but we show here the superburst observed by RXTE/PCA (Strohmayer \& Markwardt 2002). Figure \ref{fig:lc2} shows the Standard 1 mode lightcurve, which has a time resolution of 1 second, but has been binned to 10 second resolution for clarity. The best fit model agrees well with the observed decay. However, there are differences at the $\approx 10$\% level between the model and the shape of the observed lightcurve. The BeppoSAX data for the other sources have a much lower time resolution, but still allow a good constraint on the ignition depth. The fitted column depth is not very sensitive to the assumed start time of the superburst. Most important is how quickly the luminosity decays away from the peak value. For example, the count rate for 4U~1254-690 takes several hours to fall to 30\% of the peak value, whereas for 4U~1636-54 the count rate reaches 30\% of the peak after less than 2 hours. We fit only to data points that have count rates more than 10\% of the peak value. At low luminosities, the lightcurve is sensitive to how well the accretion luminosity has been subtracted, however, this uncertainty has only a small effect on the fitted column depth.

The best-fitting models have $E_{17}$ in the range $1.5$ to $2.6$. As we argued in \S 2.1, lower values of $E_{17}$ give a luminosity at early times that is smaller than observed. At the upper end of this range, Figure \ref{fig:edd} shows that the flux should exceed the Eddington flux for several minutes, inconsistent with the lack of observed photospheric radius expansion. This may indicate that the burning does not extend all the way to the surface, which our models assume, but instead stalls at a location where the thermal time to the surface is of order minutes.  More generally, our models are not valid for times less than the superburst rise time. Also, we have not fitted our models to the superburst from 4U~1820-30, which was observed by RXTE/PCA (Strohmayer \& Brown 2002). This superburst had a complex lightcurve, with an extended period of photospheric radius expansion, lasting about 1000 seconds, indicating a large energy release. More detailed 1D models which can follow the superburst rise are needed to address both of these issues.

The best-fitting column depths are in the range $0.5$--$3\times 10^{12}\ {\rm g\ cm^{-2}}$. Larger column depths closer to $10^{13}\ {\rm g\ cm^{-2}}$ are not consistent with the observed lightcurves. We have ranked the sources in Table \ref{tab:fits} in order of decreasing column depths. This ordering approximately reproduces the ordering of superbursts by their observed durations in Figure 7 of in 't Zand et al.~(2004). As pointed out in that paper, the superbursts from the rapidly accreting source GX~17+2 have low column depths, but not significantly lower than other superbursts. We find that the GX~17+2 burst has a similar ignition depth to the superbursts seen from 4U~1636-54 and Ser~X-1.

CM04 derived constraints on the ignition column depth from the observed ``quenching'' of normal Type I bursting behavior for weeks following a superburst (e.g.~Kuulkers et al.~2002). The layer continues to cool well after the superburst luminosity falls below the accretion luminosity. This residual heat flux quenches the instability of H/He burning (CB01). CM04 showed that the observed limits on the quenching timescale, although not very constraining, were at least consistent with ignition at column depths of $\approx 10^{12}$--$10^{13}\ {\rm g\ cm^{-2}}$. Our fitted values of $y_{12}$ are consistent with Figure 4 of CM04, except for KS~1731-260. The quenching timescale implies $y_{12}\gtrsim 3$ for this source, whereas our fit gives $y_{12}\approx 1$. Apart from the uncertainties associated with distance, one possibility is that the flux required to stabilize H/He burning is a factor of 3 lower than the crude estimate of CM04 (their eq.~[5]).


\section{Ignition models for superbursts}
\label{sec:ignite}

The fits to the superburst lightcurves imply ignition column depths $\approx (0.5$--$3)\times 10^{12}\ {\rm g\ cm^{-2}}$, which is accumulated in $2$--$10$ years at $0.1\ \dot m_{\rm Edd}$, or $0.6$--$3$ years at $0.3\ \dot m_{\rm Edd}$, roughly consistent with the observational constraints on recurrence times. In this section, we compare our fits to ignition models for superbursts, and use the cooling models to predict superburst properties as a function of accretion rate. In particular, we emphasize the constraints on the thermal profile of the crust and temperature of the core.

\subsection{Details of the ignition calculations and physics input}

Our ignition models follow those of Brown (2004), and are extensions of the CB01 carbon ignition models. However, we now integrate down to the crust/core interface, solving the thermal structure of the crust directly, rather than taking the outwards flux from the crust as a free parameter. Following the simplifications of Yakovlev \& Haensel (2003) and Brown (2004), we do not integrate the full structure of the star, but adopt a plane-parallel approximation, and take the gravity $g$ and gravitational redshift factor $1+z$ to be constant across the crust. Our independent variable is then the column depth $y$, where hydrostatic balance gives the relation $y=P/g$ (units of mass per unit area). We integrate the heat equation and entropy equation
\begin{equation}\label{eq:entropy}
{dF\over dy}=-\epsilon_{\rm nuc}+\epsilon_\nu
\end{equation}
\begin{equation}\label{eq:heat}
{dT\over dy}={F\over\rho K}
\end{equation}
where $K$ is the thermal conductivity, $F$ the heat flux, $T$ the temperature, $\epsilon_\nu$ the neutrino emissivity, and $\epsilon_{\rm nuc}$ the rate of heating from nuclear reactions in the crust. All quantities in equations (\ref{eq:entropy}) and (\ref{eq:heat}) are proper (local) quantities if we interpret $y$ as the rest mass column depth (e.g.~see \S 4.1 of Cumming et al.~2002). 

The flux flowing outwards from the crust $F_{\rm out}$ is a crucial parameter, since this flux heats the accumulating fuel layer. We write it in terms of the parameter $Q_b$ as $F_{\rm out}=\dot m Q_b$. Following Brown (2000), we model the heating from pycnonuclear reactions and electron captures in the crust by setting $\epsilon_{\rm nuc}=Q_{\rm nuc}\dot m/\Delta y$ to be constant over the range $\Delta y$ from $y=5.7\times 10^{15}$ to $2.2\times 10^{17}\ {\rm g\ cm^{-2}}$, where $Q_{\rm nuc}=1.4\ {\rm MeV\ per\ nucleon}$ is the total energy release in the crust. 

In the crust, the equation of state is determined by degenerate electrons and neutrons. We use the results of Mackie \& Baym (1977) to correct the neutron Fermi energy for interactions. The top of the crust is set by $\Gamma=Z^2e^2/ak_BT=175$ (Potekhin \& Chabrier 2000), where $a=(3/4\pi n_i)^{1/3}$ is the interion spacing. Since the pressure is dominated by relativistic, degenerate electrons, this translates to a column depth
\begin{equation}
y_{\rm melt}=4.5\times 10^{13}\ {\rm g\ cm^{-2}}\ \left({T_8\over 5}\right)^4\left({Z\over 26}\right)^{-8}\left({A\over 56}\right)^{4/3},
\end{equation}
where $T_8=T/10^8\ {\rm K}$. For the core, we adopt the simplified equation of state $\rho(r)=\rho_c(1-(r/R)^2)$, where $M=(8\pi/15)(\rho_cR^3)$ (Yakovlev \& Haensel 2003). This gives $\rho_c=1.7\times 10^{15}\ {\rm g\ cm^{-3}}\ (M/1.4\ M_\odot)(R/10\ {\rm km})^{-3}$. For a density at the crust/core interface of $1.6\times 10^{14}\ {\rm g\ cm^{-3}}$ (Lorentz et al.~1993), the core radius is $\approx 0.95R$. In fact, in our models, the depth of the core/crust interface from the surface is $\approx 1\ {\rm km}\approx 0.1 R$. We expect this difference to have only a small effect on our results. 

We integrate the temperature profile inwards, changing the composition from ``fuel'' to ``ash'' at a depth $y_{\rm ign}$. We set the outer boundary at $y=10^8\ {\rm g\ cm^{-2}}$, and take the temperature there to be $2\times 10^8\ {\rm K}$. We have checked that the ignition depth is not very sensitive to this choice of outer temperature\footnote{The most sensitive case is for fast cooling in the core, and $Q=100$ in the crust, for which changing the outer temperature by a factor of 2 increases the outwards flux by a factor of 2, and increases the ignition depth by 5\%.}.  We iterate to find the choice of flux at the surface that results in the flux at the base matching the core neutrino luminosity, $F_c+L_\nu(T_c)/4\pi R^2=0$. We write the core luminosity as roughly $L_\nu\approx (4\pi/3)R^3Q_\nu$, where $Q_\nu$ is the emissivity per unit volume, giving the inner boundary condition $F_c=-RQ_\nu(T_c)/3$. 

We calculate the ignition criterion for carbon according to a one-zone approximation (Fujimoto, Hanawa, \& Miyaji 1981; Fushiki \& Lamb 1987b; Cumming \& Bildsten 2000), comparing the temperature sensitivity of the heating rate at the base of the layer to the that of a local approximation to the cooling rate. Note that we calculate the temperature sensitivity of the heating rate numerically rather than assume a particular value (Brown 2004 assumed that $d\ln\epsilon_{CC}/d\ln T=26$). Although approximate, this ignition criterion agrees well with more detailed linear stability (Narayan \& Heyl 2003) and time-dependent calculations (Woosley et al.~2004) for H/He burning, and we expect it to be accurate here also (Cooper \& Narayan 2005). The carbon burning rate is given by Caughlan \& Fowler (1988) with screening from Ogata et al.~(1993)\footnote{The screening calculations of Ogata et al.~(1993) are not appropriate for the pycnonuclear regime (e.g.~Kitamura 2000; Gasques et al.~2005). However, for the temperatures $T_8>2$ that we consider in this paper, carbon burning is safely thermonuclear.}. In addition, at the ignition point we check whether the timescale for carbon depletion is longer than the accumulation time. As shown by CB01, carbon burns stably during accumulation for low accretion rates, so that the thermal instability is avoided. We show only results for which the depletion time is longer than the recurrence time. Our results compare well with those of Brown (2004). Typically, the recurrence times we find are a factor of $\lesssim 50$\% larger than those in Brown (2004), after correcting those results for gravitational redshifting.

\subsection{Neutrino cooling and crust composition and conductivity}

The main parameters in our models are the crust neutrino emissivity, the core neutrino emissivity, and the crust composition and conductivity. In the crust, we include cooling due to neutrino Bremsstrahlung according to Haensel, Kaminker, \& Yakovlev (1996) in the liquid phase, and Kaminker et al.~(1999) in the solid phase. The fitting formula given by Kaminker et al.~(1999) is for an equilibrium crust composition. To account for an accreted composition, we multiply the emissivity by a factor $R$ where $\log_{10}R=-0.2$ for $\rho<10^{11}\ {\rm g\ cm^{-3}}$, $-0.3$ for $10^{11}\ {\rm g\ cm^{-3}}<\rho<10^{13}\ {\rm g\ cm^{-3}}$, and $-0.4$ for $\rho>10^{13}\ {\rm g\ cm^{-3}}$. This closely reproduces the results for accreted matter shown in Figure 7 of Kaminker et al.~(1999), and agrees to a factor of 3 with the formula of Haensel et al.~(1996) for densities $\rho>10^{12}\ {\rm g\ cm^{-3}}$.

Most importantly, we include the possibility that the neutrons in the crust are superfluid. In this case, there is an additional neutrino cooling mechanism involving the continuous formation and breaking of Cooper pairs (Flowers, Ruderman, \& Sutherland 1976; Voskresensky \& Senatorov 1987). We use the emissivity calculated by Yakovlev, Kaminker, \& Levenfish (1999) (see eq~[\ref{Eq:PBF}]), and we take the neutron $^1S_0$ critical temperature $T_c$ as a function of density as given by the calculation of Schwenk, Friman, \& Brown (2003). We have also used the results of Ainsworth, Wambach, \& Pines (1989) (case 2 from their Fig.~3), which has a slightly different profile and a larger maximum temperature, but the results are similar and so we do not show them here. Cooper pair emission was not considered by Brown (2004) and Cooper \& Narayan (2005); however we show here that it has a dramatic effect on the crust temperature profile.

\begin{deluxetable}{llcl}
\tablewidth{0pt}
\tablecaption{Core neutrino emission\label{tab:neutrinos}}
\tablehead{\colhead{Label} &  \colhead{Type\tablenotemark{a}} & \colhead{Prefactor\tablenotemark{b}} & \colhead{Comment}\nl
\colhead{}  & \colhead{} & \colhead{$({\rm erg\ cm^{-3}\ s^{-1}})$} & \colhead{}}
\startdata
a & fast & $10^{26}$ & fast cooling\nl
b & slow & $3\times 10^{21}$ & enhanced\nl
c & slow & $10^{20}$ & mURCA\nl
d &  slow & $10^{19}$ & nn Bremsstrahlung\nl
e &  slow & $10^{17}$ & suppressed
\enddata
\tablenotetext{a}{Fast and slow cooling laws are of the form $Q_\nu=Q_f(T_c/10^9\ {\rm K})^6$ and $Q_\nu=Q_s(T_c/10^9\ {\rm K})^8$ respectively.}
\tablenotetext{b}{Either $Q_s$ or $Q_f$ for slow or fast cooling, respectively.}
\end{deluxetable}

\begin{figure}
\epsscale{1.0}
\plotone{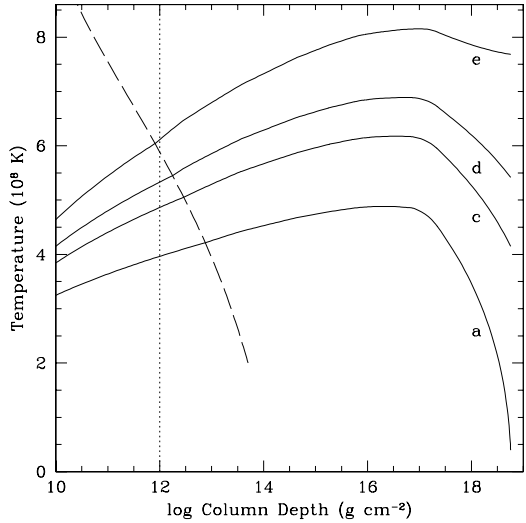}
\caption{The effect of core neutrino emissivity on superburst ignition conditions at $\dot m=0.3\ \dot m_{\rm Edd}$. We assume a disordered lattice in the crust, and do not include Cooper pairing. The accreted composition is 20\% $^{12}$C ($X_C=0.2$) and 80\% $^{56}$Fe by mass. From top to bottom, the temperature profiles are for increasing core neutrino emissivity; the letters refer to Table \ref{tab:neutrinos}. The long-dashed line shows the carbon ignition curve for $X_C=0.2$, and the vertical dotted line indicates a column depth of $10^{12}\ {\rm g\ cm^{-2}}$.
\label{fig:prof_c}}
\end{figure}

For the core neutrino emissivity, we consider the ``fast'' and ``slow'' cooling laws $Q_\nu=Q_f(T_c/10^9\ {\rm K})^6$ and $Q_\nu=Q_s(T_c/10^9\ {\rm K})^8$ (e.g.~Yakovlev \& Haensel 2003; Yakovlev \& Pethick 2004, Page et al.~2005). The ``standard'' slow cooling by modified URCA processes has $Q_s \sim 10^{20}\ {\rm erg\ cm^{-3}\ s^{-1}}$. However, if either the core protons or neutrons are superfluid, with very high values of $T_c$ ($\gg 10^9$ K), then this process is totally suppressed, leading to cooling by nucleon-nucleon Bremsstrahlung (involving the non-superfluid component). This process is roughly a factor of ten slower than modified URCA, and so we take $Q_s \sim 10^{19}\ {\rm erg\ cm^{-3}\ s^{-1}}$ in this case. If both protons and neutrons are strongly superfluid in the core,  the neutrino emission will be supressed further. To model this case, we assume that the core neutrino emission is suppressed by a further factor of 100, giving $Q_s \sim 10^{17}\ {\rm erg\ cm^{-3}\ s^{-1}}$. However, in the more reasonable case that the neutron and/or proton $T_c$ in the core are of the order of $10^9$ K there is intense neutrino emission from the Cooper pair formation, resulting in an enhanced slow cooling rate which we model by considering $Q_s \sim 3\times 10^{21}\ {\rm erg\ cm^{-3}\ s^{-1}}$ (see, e.g., Figures 20 and 21 in Page et al. 2004). Finally, we also consider a fast cooling rate with $Q_f \sim10^{26}\ {\rm erg\ cm^{-3}\ s^{-1}}$ corresponding, e.g., to the direct Urca process. These models are summarized in Table \ref{tab:neutrinos}.
The core temperature $T_c$ can be estimated in each case. For slow cooling, we find $T_c\approx 4.9\times 10^8\ {\rm K}\ ({f_{in}^{1/8}/Q_{s,20}^{1/8}})\left({\dot m/\dot m_{\rm Edd}}\right)^{1/8}$ and fast cooling $T_c\approx 5.0\times 10^7\ {\rm K}\ ({f_{in}^{1/6}/Q_{f,26}^{1/6}})\left({\dot m/\dot m_{\rm Edd}}\right)^{1/6}$ where $f_{in}$ is the fraction of heat released in the crust that is conducted into the core. 

For the composition of the crust, we use the composition calculated by either Haensel \& Zdunik (1990) or Haensel \& Zdunik (2003). The difference between these two calculations is the nucleus assumed to be present at low densities, either $^{56}$Fe (Haensel \& Zdunik 1990), or a heavy nucleus $^{106}$Pd ($Z=46$) (Haensel \& Zdunik 2003), as would be appropriate if  rp-process hydrogen burning is able to run to its endpoint (Schatz et al.~2001). We calculate results for these two cases to illustrate the variation expected from changes in composition. For the conductivity, we consider two cases. The first is a ``disordered'' crust, for which we take the conductivity to be that of a liquid phase, in the second case, we calculate the contributions from phonons (Baiko \& Yakovlev 1996) and electron-impurity scattering (Itoh \& Kohyama 1993), taking the impurity parameter $Q=100$ (see Itoh \& Kohyama 1993 for a definition of the impurity parameter, written as $\langle(\Delta Z)^2\rangle$ in their notation). Note that a crust with $Q=100$ is very impure. However, we do not consider smaller values of the impurity parameter because as we will show they would not agree with observed X-ray burst properties.

\begin{figure}
\epsscale{1.0}
\plotone{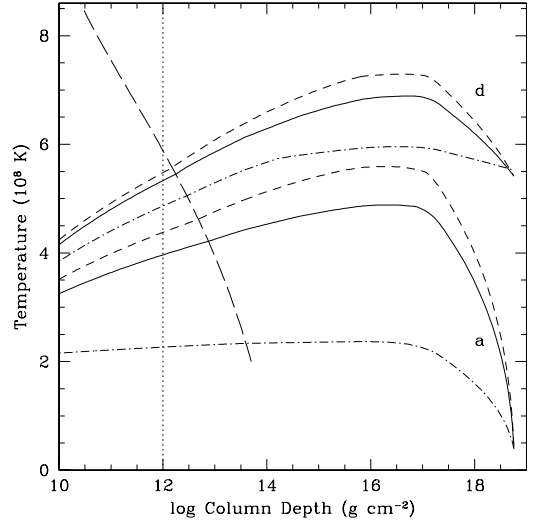}
\caption{The effect of crust composition and conductivity on superburst ignition conditions. Temperature profiles for superburst ignition models at $\dot m=0.3\ \dot m_{\rm Edd}$. We show two cases of core neutrino emissivity: slow cooling with $Q_s=10^{19}\ {\rm erg\ cm^{-3}\ s^{-1}}$ and fast cooling with $Q_f=10^{26}\  {\rm erg\ cm^{-3}\ s^{-1}}$. Solid lines are for a composition of $^{56}$Fe and a disordered lattice. Short-dashed lines have a heavier composition ($A=106, Z=46$), and dot-dashed lines are for a larger thermal conductivity ($Q=100$). The long-dashed line shows the carbon ignition curve for $X_C=0.2$, and the vertical dotted line indicates a column depth of $10^{12}\ {\rm g\ cm^{-2}}$.\label{fig:prof_c2}}
\end{figure}

\begin{figure}
\epsscale{1.0}
\plotone{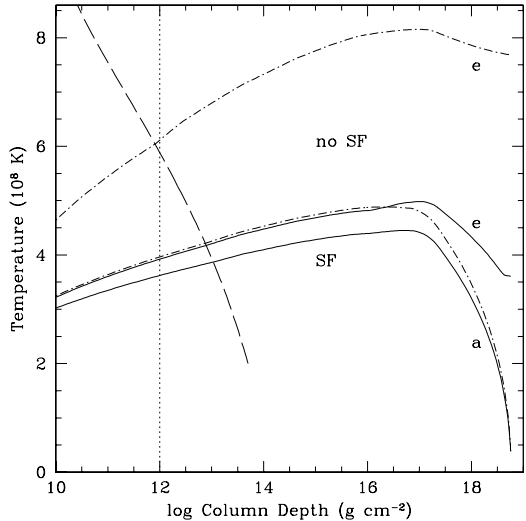}
\caption{The effect of neutrino cooling by Cooper pairs on superburst ignition conditions. For two different core neutrino emissivities, we show temperature profiles with (solid) and without (dot-dashed) neutrino cooling by Cooper pairs. These models are for $\dot m=0.3\ \dot m_{\rm Edd}$ and have $X_C=0.2$. The long-dashed line shows the carbon ignition curve for $X_C=0.2$, and the vertical dotted line indicates a column depth of $10^{12}\ {\rm g\ cm^{-2}}$.\label{fig:prof_c3}}
\end{figure}

\subsection{Ignition conditions at a fixed accretion rate}

We first calculate ignition conditions  for carbon at $\dot m=0.3\ \dot m_{\rm Edd}$, for different values of neutrino emission and crust properties. In order to illustrate the effects of different parameters, we start by assuming that the neutrons in the crust are normal (no cooling due to Cooper pair neutrinos). For this case, Figure \ref{fig:prof_c} shows the effect on the temperature profile of varying the core neutrino emissivity. Less efficient neutrino emission leads to a higher core temperature and a greater fraction of the energy released in the crust is emitted through the surface, heating the carbon layer.  At this accretion rate, standard slow cooling in the core results in recurrence times $>3$ years, longer than inferred from observations. Some suppression of the modified URCA rate is necessary to bring the predicted and observed recurrence times into agreement.

Figure \ref{fig:prof_c2} illustrates the effect of the crust composition and conductivity on the temperature profile for fast and slow cooling in the core. At $\dot m=0.3\ \dot m_{\rm Edd}$, the model with $Q=100$ and fast core neutrino emission gives superburst recurrence times that are much longer than observed. In general, the change in ignition depth with composition is much smaller than the change in ignition depth with other model parameters. This is because a heavier composition decreases the thermal conductivity, but also results in less outwards flux from the crust, as pointed out by Brown (2004) and Cooper \& Narayan (2005). In Appendix B we give a simple analytic argument to understand this.

We now include neutron superfluidity in the crust. Figure \ref{fig:prof_c3} shows the dramatic effect of the extra cooling from Cooper pair neutrino emission. We show profiles for either fast cooling in the core, or highly suppressed core cooling, with and without Cooper pair cooling in the crust. When Cooper pairing of neutrons is included, the temperature is limited by neutrino losses to a value $T\lesssim 5\times 10^8\ {\rm K}$. This is true even for highly suppressed neutrino cooling in the core; in this case, most of the energy release in the crust leaves as neutrino emission from within the crust itself. In Appendix B, we show how to understand this limiting temperature analytically by balancing the heating and cooling rates. 

\begin{figure}
\epsscale{1.0}
\plotone{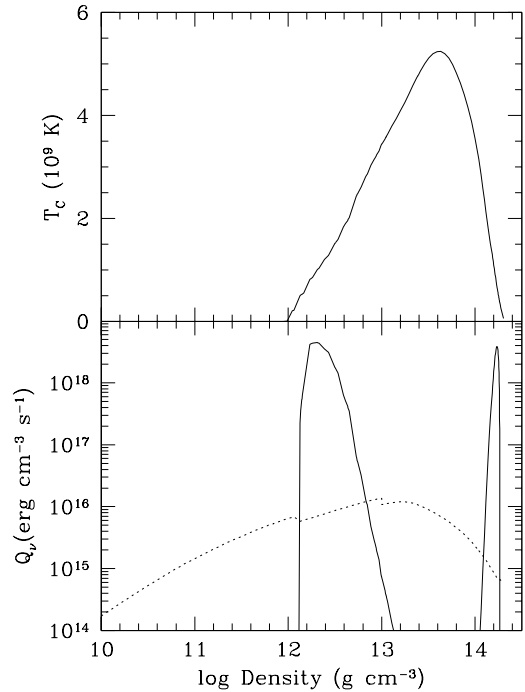}
\caption{The critical temperature for neutron superfluidity in the crust, according to Schwenk et al.~2003, and an example of the neutrino emissivity as a function of depth from Cooper pairing (solid lines) and electron bremsstrahlung (dotted line), for model ``e'' with Cooper pairing shown in Figure \ref{fig:prof_c3}. The low density peak in the Cooper pair emissivity corresponds to column depths in the range $10^{16}$--$10^{17}\ {\rm g\ cm^{-2}}$, the peak at higher densities corresponds to column depths $\approx 3\times 10^{18}\ {\rm g\ cm^{-2}}$ near the base of the crust.
\label{fig:profq}}
\end{figure}

Figure \ref{fig:prof_c3} shows that if Cooper pairing is important in the crust, the superburst recurrence times should be long $\approx 10\ {\rm years}$, and insensitive to core neutrino emissivity. In Figure \ref{fig:profq}, we show the critical temperature for the neutron superfluid, and  the neutrino emissivity from Cooper pairing and electron Bremsstrahlung for comparison. The Cooper pair emission is concentrated in two regions where $T\sim T_c$, as discussed by Yakovlev et al.~(1999), and is therefore sensitive to the behavior of $T_c$ close to the superfluid threshold, which is uncertain. We address these uncertainties in Appendix C.   However, we find that as long as the critical temperature increases from zero to large values, crossing the crust temperature, the process is important. For example, we have tried modelling the $T_c$ profile as log-Gaussian in density, and varying the central density and width, but have not been able to significantly reduce the Cooper pair neutrino emission. The peak in emission at lower densities has the largest effect, since  this extra cooling occurs at the location of the energy release in the crust (close to neutron drip). The peak in emissivity near the core boundary has a smaller effect, equivalent to an extra core neutrino emission.

\begin{figure}
\epsscale{1.1}
\plotone{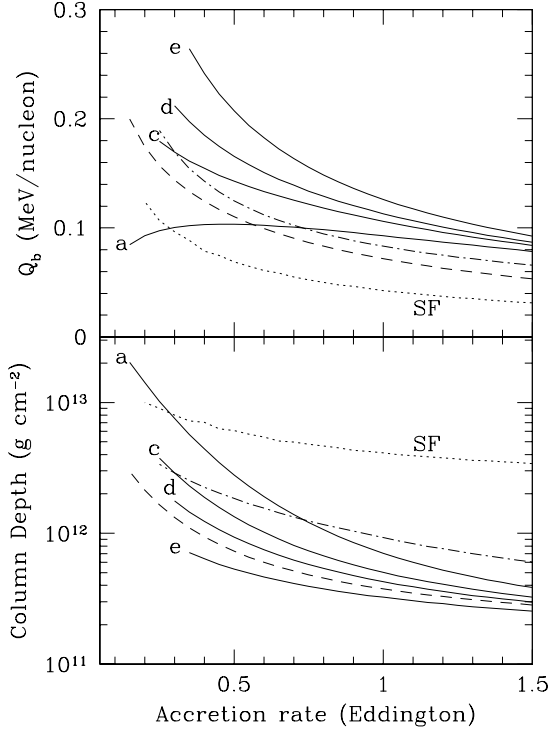}
\caption{Flux from the crust heating the fuel layer and ignition column depth as a function of accretion rate. The solid curves show results for a disordered crust, and a composition of $X_C=0.2$ and $^{56}$Fe, for the four different core neutrino emissivities of Figure \ref{fig:prof_c}. More efficient core neutrino emission gives a lower flux from the crust $Q_b$, and a larger ignition column depth. Other curves show variations on the ``d'' model. The short-dashed curve is for a heavier composition ($A=106, Z=46$); the dot-dashed curve is a higher crust conductivity ($Q=100$); the dotted curve includes Cooper pairing in the crust (labelled ''SF''). The results with Cooper pairing are not very sensitive to the core neutrino emissivity. At low accretion rates, the curves are terminated at the accretion rate where the carbon begins to burn stably (defined as the point where depletion time for carbon equals the recurrence time). 
\label{fig:qy}}
\end{figure}

\begin{figure}
\epsscale{1.1}
\plotone{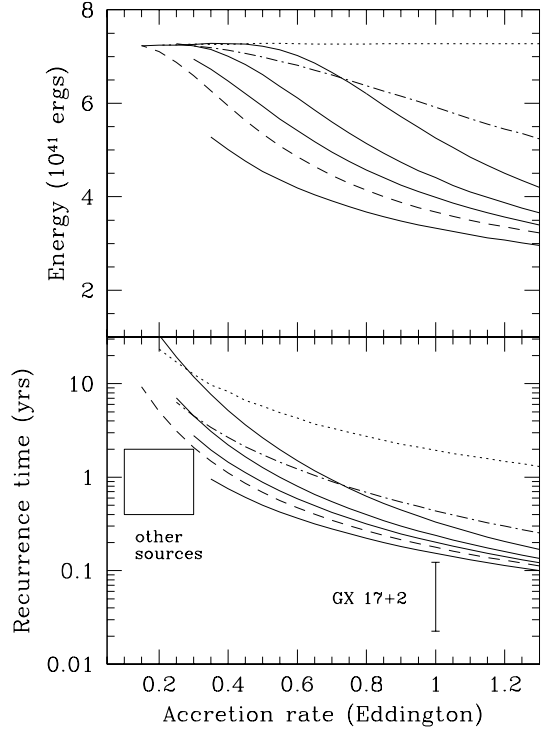}
\caption{For the models shown in Figure \ref{fig:qy}, we plot the energy released from the surface in the first 6 hours following ignition (as calculated using the cooling models of \S 2), and the recurrence time. The errorbars indicate the estimated recurrence times and accretion rates for most superburst sources and for the near-Eddington accretor GX~17+2. \label{fig:obs}}
\end{figure}

\subsection{Variation with accretion rate and comparison to observations}

CB01 showed that the ignition conditions are very sensitive to accretion rate, and so a natural question is how much the results of \S 3.3 depend on the assumed accretion rate. We explore this dependence here. The accretion rate in the superburst sources is believed to lie in the narrow range $0.1$--$0.3$ Eddington, but there is some uncertainty in these estimates due to for example uncertainty in the relation between accretion rate and X-ray luminosity, and distance uncertainties. Figure \ref{fig:qy} shows the ignition column depth and the energy per gram released in the crust that flows outwards $Q_b$ as a function of accretion rate for different models. 

For normal crust neutrons, the constraint on core neutrino emission can be relaxed if the accretion rate is in fact larger than inferred from the observed X-ray luminosity. For example at $\dot m\approx 0.5\ \dot m_{\rm Edd}$, standard slow cooling from modified URCA reactions explains the observed ignition columns. However, with Cooper pair cooling in the crust (dotted curve in Fig.~\ref{fig:qy}), the ignition columns remain well above $10^{12}\ {\rm g\ cm^{-2}}$ even at accretion rates $\approx \dot m_{\rm Edd}$. This implies that some extra heating of the carbon layer must occur that is not included in our model.

Another important point is that with $X_C=0.2$, unstable ignition requires $\dot m\gtrsim 0.3\ \dot m_{\rm Edd}$, because at lower rates the carbon burns stably (the curves in Figure \ref{fig:qy} terminate on the left where the carbon begins to burn stably). A carbon fraction $X_C\gtrsim 0.2$ is consistent with the results of our lightcurve fits, which gave $E_{17}\gtrsim 2$ in most cases.

Predictions for the observable quantities recurrence time and superburst energy are shown in Figure \ref{fig:obs}. We indicate the observed constraints on recurrence time for superburst sources, and separately for the rapidly accreting source GX~17+2. Only the models with low neutrino emissivity in the crust and core come close to the observed values at the estimated accretion rates. The accretion rate for GX~17+2 is quite uncertain; we adopt a value of $\dot m_{\rm Edd}$ for this source. For a given ignition depth and carbon fraction, we use our cooling models to predict the superburst energy, which we take to be the energy released in the surface in the first 6 hours. Note that the energies are close to $10^{42}\ {\rm ergs}$ in Figure \ref{fig:obs} because we take $X_C=0.2$ in these models, the energy would be significantly smaller if $X_C=0.1$. The behavior of the superburst energy with accretion rate is in general not constraining. For long recurrence times $\gtrsim 1$ year, the superburst energy saturates at $\approx 10^{42}\ {\rm erg}$ because of the effects of neutrino cooling and inwards conduction of heat, as discussed in \S 2.

The overall conclusion is that to achieve ignition at column depths implied by our fits requires inefficient neutrino cooling from the core and the crust, and accretion rates larger than inferred from the X-ray luminosity. Even with normal neutrons and core neutrino emission that is less efficient than modified URCA, it is difficult to reproduce the observed superburst recurrence times and column depths at accretion rates as low as $0.1\ \dot m_{\rm Edd}$. Neutrino cooling from the crust due to Cooper pair formation results in ignition depths that are too large even for accretion rates near Eddington. When this process is included, our models cannot reproduce the observed superburst recurrence times. In addition, carbon fractions of $\gtrsim 0.2$ are required to avoid stable burning of the carbon and achieve unstable ignition at accretion rates of $0.3\ \dot m_{\rm Edd}$.


\section{Ignition models for pure helium bursts}
\label{sec:helium}

We now consider the constraints that come from pure helium accretors. Pure helium bursts are interesting because they can occur at a wide range of accretion rates. Two sources in particular are thought to be accreting pure helium (perhaps with a small amount of hydrogen; Podsiadlowski, Rappaport, \& Pfahl 2002), with accretion rates different by an order of magnitude. The ultracompact binary 4U~1820-30 has an orbital period of only 11.4 minutes (Stella, Priedhorsky, \& White 1987), implying a hydrogen-poor companion. This source shows frequent and regular Type I X-ray bursts whose properties are consistent with accretion of pure helium at rates close to  $\dot M\approx 0.1$--$0.2\ \dot M_{\rm Edd}$ as inferred from the X-ray luminosity at the time when bursts are seen (Bildsten 1995; Cumming 2003a). The persistent X-ray source 2S~0918-549 is suspected also to be an ultracompact binary because of its low optical to X-ray flux ratio, and lack of hydrogen lines in its optical spectrum (Juett et al.~2001; Nelemans et al.~2004). Recently, a long duration X-ray burst was observed from this source whose properties can be explained by accretion of pure helium at the observed rate of $\dot M\approx 0.01\ \dot M_{\rm Edd}$ (in~'t Zand et al.~2005).

\begin{figure}
\epsscale{1.1}
\plotone{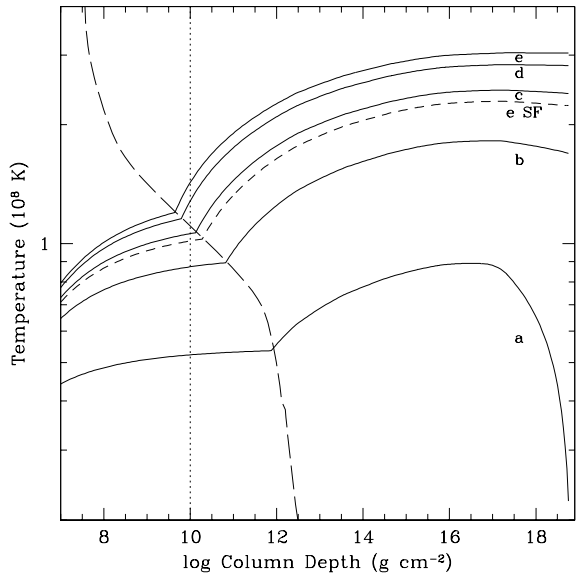}
\caption{Temperature profiles for pure helium ignition models at $\dot m=0.01\ \dot m_{\rm Edd}$, with a disordered lattice. We show five examples of core neutrino emissivity, a to e from Table 2. The dot-dashed curve is model e including Cooper pair emission from the crust. The long-dashed curve is the triple alpha ignition curve. The dotted line marks a column depth of $10^{10}\ {\rm g\ cm^{-2}}$, as inferred for the long burst from 2S~0918-549 observed by in 't Zand et al.~(2005).\label{fig:prof_he}}
\end{figure}

\begin{figure}
\epsscale{1.0}
\plotone{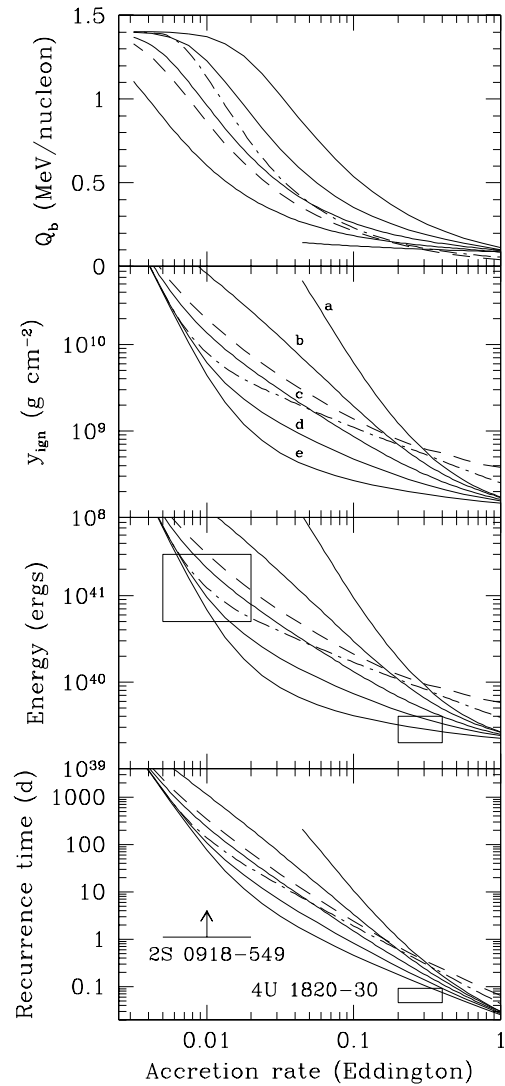}
\caption{The outwards heat flux $Q_b$, ignition column depth $y_{\rm ign}$, and predicted energy and recurrence times for pure helium flashes, for the same models as Figure \ref{fig:prof_he}. In addition, we show a model with $Q=100$ in the crust for core emission ``c'' (dot-dashed curve), and a model with Cooper pairs included for core emission ''e'' (dashed line).
\label{fig:all_he}}
\end{figure}

\begin{figure}
\epsscale{1.1}
\plotone{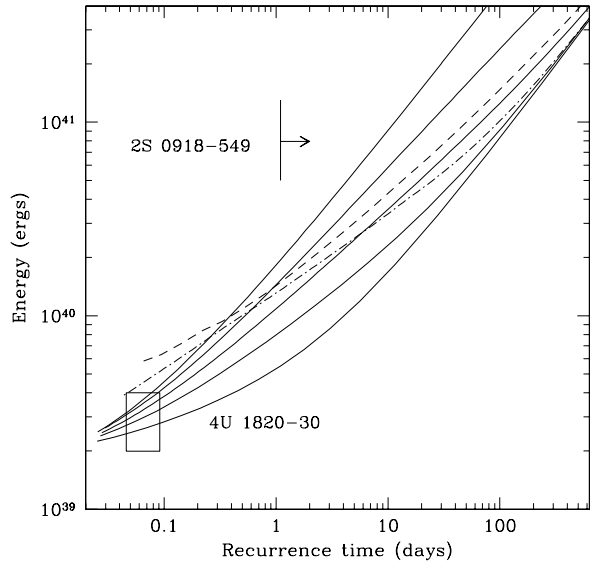}
\caption{Predicted burst energy against recurrence time for pure helium flashes. For comparison, we show observed burst properties for 4U~1820-30, and 2S~0918-549. The solid curves are for a disordered crust with different core neutrino emissivities, the dot-dashed curve is for standard slow cooling with $Q=100$ in the crust; the dashed curve is for Cooper pairing in the crust and core emission ''e''.\label{fig:et_he}}
\end{figure}

We have calculated ignition conditions for pure helium bursts as a function of accretion rate, crust composition, and core neutrino emissivity. The calculations follow those for superburst ignition in \S 3, except that the accumulated fuel is pure helium, and the nuclear burning rate is given by the triple alpha rate from Fushiki \& Lamb (1987a). We start integrating at a column depth of $10^3\ {\rm g\ cm^{-3}}$, and set the temperature there proportional to $F_b^{1/4}$, although the solutions are not very sensitive to the outer temperature in most cases. 

We first consider accretion at a rate $\dot m=0.01\ \dot m_{\rm Edd}$ appropriate for 2S~0918-549. Temperature profiles for this accretion rate are shown in Figure \ref{fig:prof_he}. We show profiles for the five different core neutrino emissivities in Table 2, and for core emissivity ``e'' but including Cooper pair emission. At $\dot m\approx 0.01\ \dot m_{\rm Edd}$, Cooper pair emission limits the crust temperature to $T\approx 2.3\times 10^8\ {\rm K}$ (see eq.~[\ref{eq:cooperT}] in Appendix B for an analytic estimate). The long burst from 2S~0918-549 (in 't Zand et al.~2005) had an energy of $\approx 10^{41}\ {\rm ergs}$, implying an ignition column depth of $y\approx 10^{10}\ {\rm g\ cm^{-2}}$ for an energy release of $\approx 10^{18}\ {\rm erg\ g^{-1}}$ appropriate for helium burning to heavy elements. This column depth is also consistent with the burst lightcurve, which is well-fitted by a cooling model based on this column depth (in 't Zand et al.~2005).  Figure \ref{fig:prof_he} shows that ignition at $y\approx 10^{10}\ {\rm g\ cm^{-2}}$ requires that the core neutrino emissivity not be more efficient than modified URCA. Enhanced slow cooling (model b), e.g.~by Cooper pairing in the core, or fast core cooling (model a) lead to ignition at columns of $\gtrsim 10^{11}\ {\rm g\ cm^{-2}}$.

This conclusion is sensitive to the assumed accretion rate. However, models d and e require substantial increases in the accretion rate to obtain burst energies $\approx 10^{41} \ {\rm ergs}$, by factors of 5--10 above the inferred rate of $0.01\ \dot M_{\rm Edd}$. Figure \ref{fig:all_he} shows the variation in the heat flux from the crust $Q_b$, ignition depth, and burst energy and recurrence time with accretion rate. The ignition conditions at low accretion rates are not very sensitive to the crust composition, and depend mostly on core neutrino emission. For comparison with the disordered crust models, a crust with $Q=100$ and standard modified URCA slow cooling in the core is shown as the dashed curve in Figure \ref{fig:all_he}.

4U~1820-30 accretes at a rate comparable to the superburst sources, $\dot m\approx 0.2\ \dot m_{\rm Edd}$.  The burst properties observed in 4U~1820-30 are shown in Figure \ref{fig:all_he}. This source undergoes periodic variations in accretion rate, with bursts being seen in the low state when the accretion rate is a factor of two or more below the time-averaged rate. Because the crust temperature profile is set by the time-averaged rate, we correct for this by plotting the burst properties at the time-averaged rate, and decreasing the recurrence time by a factor of two. The recurrence time is again better explained if the neutrino emission is inefficient in the core and crust. Whereas Cooper pair neutrino emission does not significantly affect the recurrence times at low accretion rates, it does make a difference at the higher accretion rates appropriate for 4U~1820-30 because of the larger crust temperatures. Our results are consistent with the previous analysis of Cumming (2003a), who noted that to achieve ignition for pure helium at the observed rates required a flux from the crust of $Q_b\approx 0.4$ MeV per nucleon. An estimate of the residual heat released between bursts was not enough to account for this extra flux; inefficient core neutrino emission offers a new explanation. Another complication for this source is that the accreted material may contain a small amount of hydrogen, which can significantly heat the accumulating layer and shorten the recurrence time (Cumming 2003a).

An accurate  measurement of recurrence time for the long helium burst would impose further constraints. In Figure \ref{fig:et_he} we plot burst energy against recurrence time for the different models. This plot shows clearly that the long duration bursts at low accretion rates are most sensitive to core neutrino emission, whereas short bursts at higher accretion rates are most sensitive to crust composition. The two systems  4U~1820-30 and 2S~0918-549 therefore offer complementary constraints on the interior model. Unfortunately, the recurrence time for the 2S~0918-549  burst is not well constrained (in't Zand et al.~2005). In Figure \ref{fig:et_he}, we show the observed lower limit of 1.1 days.

To summarize, the conclusions for pure helium bursts are the same as for superbursts. Enhanced core neutrino emission relative to modified URCA leads to larger ignition columns, energies, and recurrence times than observed for  2S~0918-549. Cooper pairing in the crust leads to much larger recurrence times and energies than observed for 4U~1820-30. The constraints again depend on the assumed accretion rate, and can be relaxed if the accretion rate is larger than inferred from the X-ray luminosity by factors of $\gtrsim 2$.


\section{Summary and Discussion}
\label{sec:conc}

In this paper, we have compared models of carbon and helium ignition on accreting neutron stars to observations of long duration X-ray bursts. In particular, we have investigated the effect of the thermal profile of the crust and core on the ignition conditions, and how well the ignition conditions reproduce the observed burst properties.  We have improved on the earlier work of Brown (2004) and Cooper \& Narayan (2005) by (i) using cooling models of superbursts to predict observational properties, and then using these lightcurves to provide an independent constraint on ignition depth and energetics, (ii) including neutrino cooling in the inner crust due to Cooper pairing of neutrons, and (iii) considering pure helium accretion in addition to carbon.

\subsection{Superbursts}

We applied the cooling models for superbursts calculated by Cumming \& Macbeth (2004) to observed superburst lightcurves. Despite the large uncertainties in distance, we find that the energy release and ignition column depths are quite well constrained, with the best fitting models giving $E_{17}\approx 2$ and $y_{12}=0.5$--$3$. Lower values of $E_{17}$ give a luminosity at early times that is lower than observed, or equivalently, total superburst energies much smaller than the observed energies of $\approx 10^{42}\ {\rm ergs}$. An upper limit on $E_{17}$ comes from the lack of photospheric radius expansion observed in most superbursts. Large values of $E_{17}\gtrsim 2$ lead to extended periods of super-Eddington luminosities (durations of minutes and longer), inconsistent with observations. The column depth is determined by the rate at which the luminosity falls in the tail of the superburst. For example, the range of fitted values $y_{12}=0.5$--$3$ goes from 4U~1636-54 at one end to 4U~1254-690 at the other. The superburst from 4U~1254-690 is the longest observed, taking 7 hours to fall to a luminosity 30\% of the peak value, whereas the superburst from 4U~1636-54 had a much shorter duration, falling to 30\% of the peak luminosity after roughly 1.5 hours. In our models, large ignition column depths of $\approx 10^{13}\ {\rm g\ cm^{-2}}$ lead to much slower decays than observed.

The fitted column depths are roughly consistent with observational constraints on recurrence times. At the moment, only 4U~1636-54 and GX~17+2 have shown multiple superbursts. For 4U~1636-54 we derived a column depth of $y_{12}\approx 0.5$ by fitting the superburst lightcurve. The time between the previous superburst and the superburst observed by RXTE/PCA was $1.75$ years (Kuulkers et al.~2004). Taking this to be the recurrence time and combining with the fitted ignition column gives an accretion rate $\dot m=0.10\ \dot m_{\rm Edd}$, exactly as inferred from the persistent X-ray luminosity. For GX~17+2, the mean recurrence time is 30 days (in 't Zand et al.~2004), giving an accreted column depth of $2\times 10^{11}\ {\rm g\ cm^{-2}}$ for accretion at $\dot m=\dot m_{\rm Edd}$. The fitted column depth for burst 2 from this source is $6\times 10^{11}\ {\rm g\ cm^{-2}}$, a factor of 3 larger. The general constraints on superburst recurrence times are $0.4$--$2$ years (in't Zand et al.~2003), which gives column depths $y\approx 1$--$5\times 10^{11}\ {\rm g\ cm^{-2}}\ (\dot m/0.1\ \dot m_{\rm Edd})$.

\begin{figure}
\epsscale{1.0}
\plotone{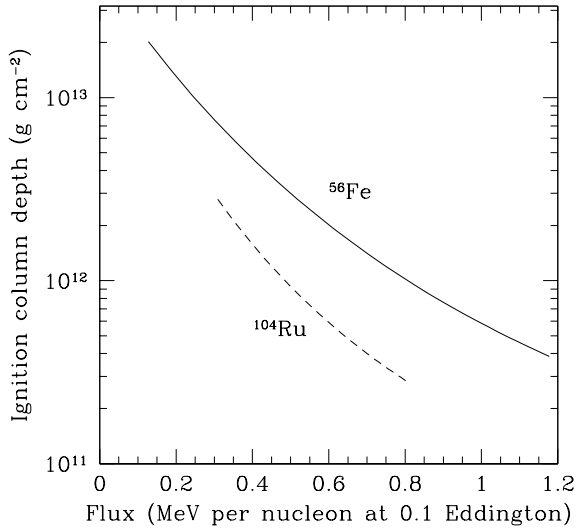}
\caption{Ignition column depth for carbon as a function of the flux from below heating the accumulating fuel layer. We write the flux as the equivalent energy per nucleon at 0.1 of the Eddington accretion rate. The solid curve are models with iron as the heavy element, the dashed curve is for a heavy composition $A=104$, $Z=44$. The ignition column is not very sensitive to carbon abundance; for this Figure we assume $X_C=0.2$.
\label{fig:baseflux}}
\end{figure}

How do the fitted column depths compare to carbon ignition models for superbursts? The most striking result is that to achieve ignition at the inferred column depths for accretion rates thought to be appropriate for these sources, $0.1$--$0.3\ \dot M_{\rm Edd}$, requires adjusting each parameter to maximize the flux emerging from the crust. To illustrate this, Figure \ref{fig:baseflux} shows the ignition column depth as a function of the base flux. We find that the flux required for ignition\footnote{We give the result here for a heavy element of $^{56}$Fe. The argument for a heavier composition is similar because as Brown (2004) pointed out, the flux emerging from the crust is smaller, compensating for the fact that a smaller flux is needed for ignition at a particular column depth. Nonetheless, a heavier composition does give a smaller ignition column depth, but the crust and core properties are more important parameters.} at $y=10^{12}\ {\rm g\ cm^{-2}}$ is $Q_b\approx 0.25\ (\dot m/0.3\ \dot m_{\rm Edd})^{-1}$. Figure \ref{fig:qy} shows that this value of flux requires that (i) crust cooling by Cooper pairs is not active, (ii) the core neutrino emission is significantly reduced below modified URCA, and (iii) the crust conductivity should be ``disordered'' so that it has a low thermal conductivity. We now discuss the likelihood of satsifying each of these requirements.

Most important is the effect of Cooper pairs in the crust. The emissivity due to this process is large wherever the temperature of the crust is close to the critical temperature for superfluidity $T_c$, because this allows the efficient formation and breaking of Cooper pairs (Yakovlev et al.~1999). This happens near the base of the crust, but most importantly close to neutron drip where most of the nuclear energy release in the crust occurs. The effect is to limit the temperature of the crust to be $\lesssim 4.4\times 10^8\ {\rm K}\ (\dot m/\dot m_{\rm Edd})^{1/7}$ (eq.~[\ref{eq:cooperT}]). In Appendix C, we discuss the uncertainties in this cooling mechanism. The emissivity depends on the $T_c$ profile with density, however all models that we have tried give substantial cooling rates from this process, resulting in ignition columns for superbursts $\gtrsim 4\times 10^{12}\ {\rm g\ cm^{-2}}$ for accretion rates less than the Eddington rate. Some extra heating of the accumulating carbon layer is needed to explain observed superburst properties if Cooper pair cooling is active.

\begin{figure}
\epsscale{1.0}
\plotone{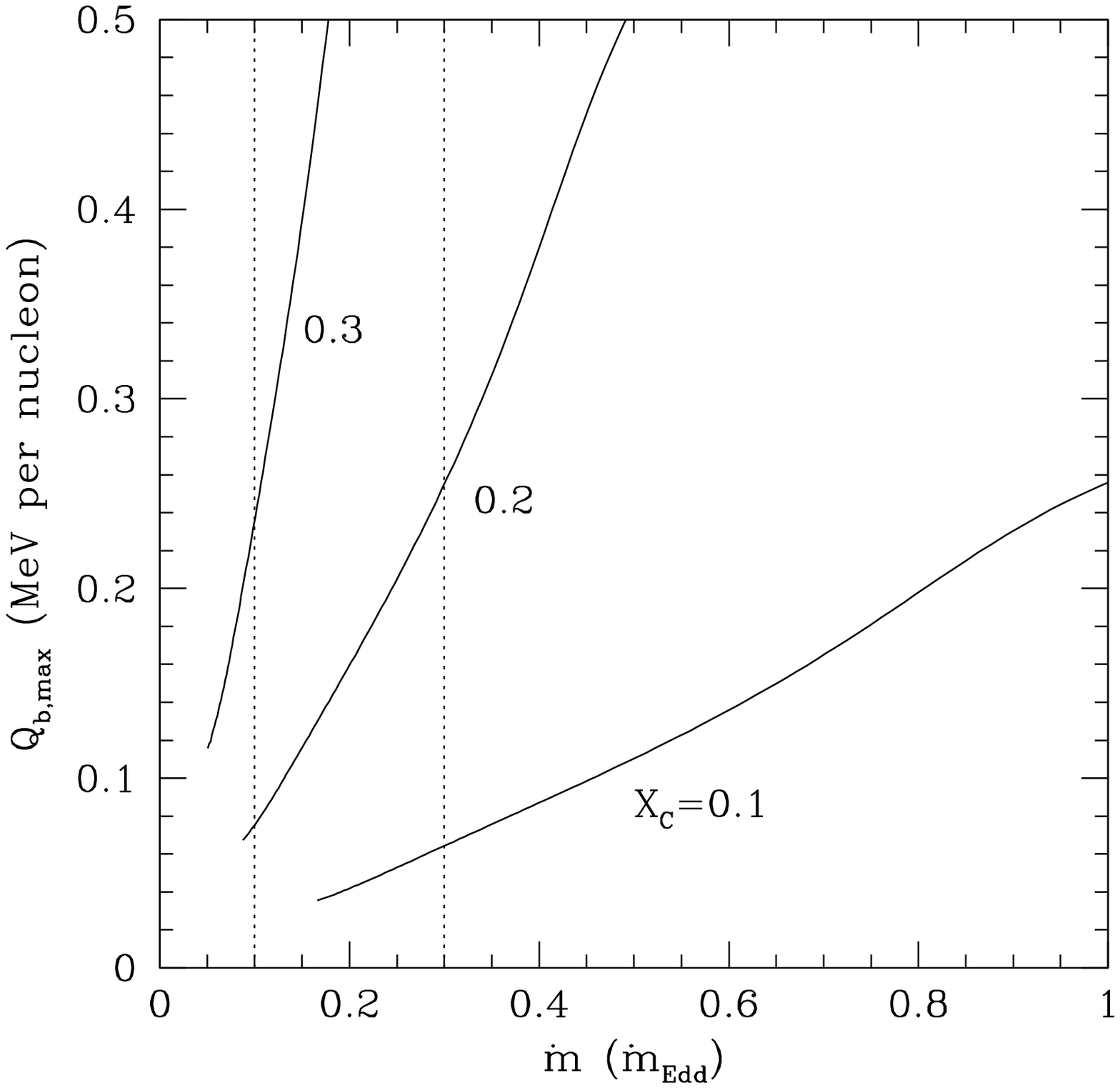}
\plotone{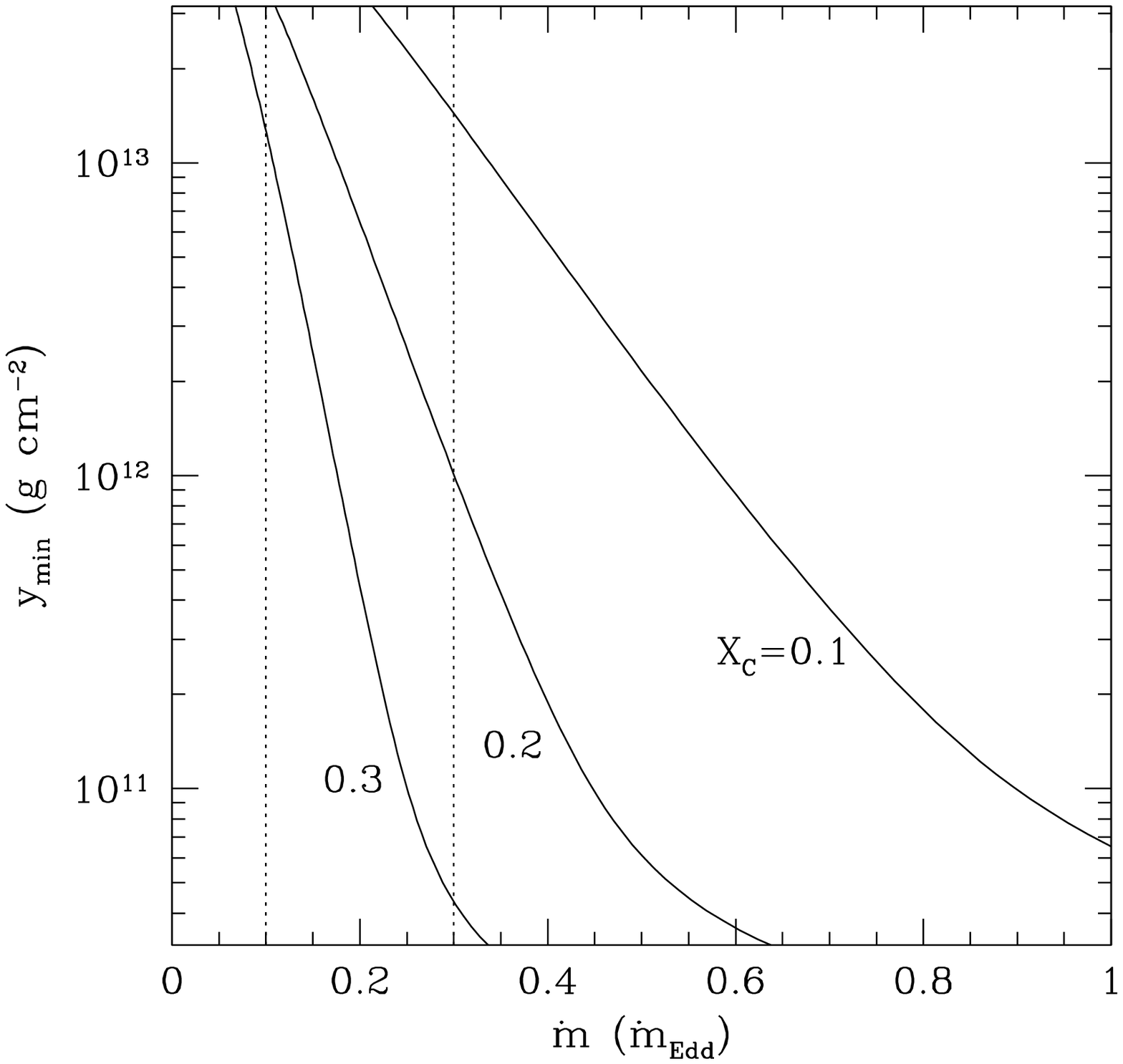}
\caption{Maximum outwards flux and minimum column depth for unstable carbon ignition for different carbon fractions. At a given accretion rate, a larger flux than indicated in the upper panel results in stable rather than unstable burning of carbon. This translates into the minimum possible column depth for unstable carbon ignition shown in the lower panel. The dotted lines indicate the range of accretion rates for most superburst sources, $0.1$--$0.3\ \dot m_{\rm Edd}$.\label{fig:ymin}}
\end{figure}

\begin{figure}
\epsscale{1.0}
\plotone{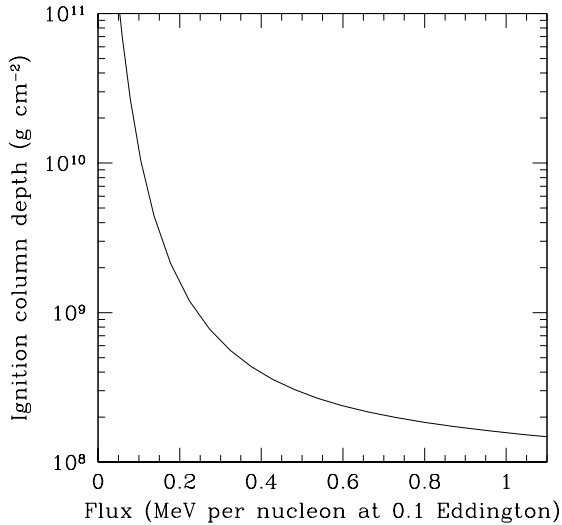}
\caption{Ignition column depth for pure helium as a function of the flux from below heating the accumulating fuel layer. We write the flux as the equivalent energy per nucleon at 0.1 of the Eddington accretion rate.
\label{fig:basefluxhe}}
\end{figure}

Suppression of the core neutrino emissivity below the modified URCA rate
will occur if either the protons or neutrons are superfluid in the core. 
However layers whose temperature $T \leq T_c$ will copiously emit neutrinos 
through
the Cooper pair process and hence reduction of the core neutrino luminosity
requires critical temperatures for neutrons and/or protons wich satisfy either 
$T_c<T$ or $T_c \gg T$ {\em everywhere} in the core.
If both neutrons and protons are paired in the whole core with $T_c$'s 
well above $10^9$ K everywhere, then only minor neutrino emission processes 
may be active and the model (e) of Table \ref{tab:neutrinos} would apply.
Nevertheless, in this case the star's specific heat would be very reduced
leading to rapid cooling of isolated cooling neutron stars of age around 10$^5$ years, in conflict with observations
(see, e.g., Gusakov et al.~2004).
Yakovlev \& Pethick (2004) argue that the best calculations of transition 
temperatures imply that the protons are superfluid, but neutrons not. 
In the case of proton superconductivity, and assuming $T_c \gg 10^9$ K,
the core neutrino emissivity would be set by neutron-neutron Bremsstrahlung, 
roughly an order of magnitude below the modified URCA rate, 
corresponding to model (d) of Table \ref{tab:neutrinos}.
However, all calculations of proton gaps show that $T_c$ decreases and eventually
vanishes at relatively low density (see, e.g., Figure 9 in Page et al. 2004)
and hence strong proton superconductivity, to avoid the proton $^1S_0$ Cooper
pair neutrinos, probably requires a very low mass star.
Moreover, considering that all superbursters are heavily accreting,
and probably also very old, some sources may be massive stars in which case, 
due to the resulting high central density, numerous fast neutrino processes
of the direct URCA family are possible (see, e.g., Page et al. 2005 for a review).

Finally, the crust conductivity is expected to be low in accreting neutron stars, because hydrogen and helium burning produces a complex mixture of heavy elements (Schatz et al.~1999). Our results suggest a completely disordered crust with a thermal conductivity essentially equivalent to that of the liquid state. Even a very impure crust with impurity parameter $Q=100$ does not fit the data as well as a completely disordered crust. Interestingly, recent work by Jones (2001, 2004a, 2004b) concludes that the same is likely to be true for the original crust before accretion starts.

It is very important to stress that conclusions about the interior thermal properties are sensitive to the choice of local accretion rate. The constraints on core emissivity can be relaxed by an increase in accretion rate by factors of 2 or 3. This is a possibility since the  relation between X-ray luminosity and accretion rate is uncertain, and the accreted material may cover only part of the stellar surface (Bildsten 2000). However, we stress that if Cooper pair cooling operates, we cannot reproduce observed superburst properties even for accretion rates near the Eddington rate. 

The carbon ignition models also show that a large carbon fraction $X_C\gtrsim 0.2$ is needed if conditions for the thermal instability are to be achieved before the carbon stably burns away. This is illustrated in Figure \ref{fig:ymin}, which shows the maximum value of $Q_b$ that allows stable burning as a function of $\dot m$ and $X_C$. For larger base fluxes, the carbon burns stably. This limit on base flux translates into a lower limit to the ignition column depth at particular accretion rate, shown in the lower panel of Figure \ref{fig:ymin}.  Ignition at column depths near $10^{12}\ {\rm g\ cm^{-2}}$ at the observed accretion rates for superburst sources requires $X_C>0.2$. This conclusion is consistent with the lightcurve fits, which imply $E_{17}\approx 2$, if carbon burning releases $\approx 10^{18} X_C\ {\rm erg\ g^{-1}}$ as expected for carbon burning to iron group. There is no need for additional energy release, for example from photodisintegration of heavy elements (Schatz, Bildsten, \& Cumming 2003a). The large inferred carbon fraction is an important constraint on models of rp-process hydrogen and helium burning. Current models suggest that stable burning of hydrogen and helium most likely plays a role in producing the fuel (in 't Zand et al.~2003; Schatz et al.~2003b), although further work is needed.

\subsection{Pure helium bursts}

Pure helium bursts are interesting because they probe lower ignition masses and densities than superbursts, and can occur at a wider range of accretion rates. The likely ultracompact binary 2S~0918-549 is a persistent source as far as is known, and accretes at a rate $\approx 0.01\ \dot M_{\rm Edd}$. By considering the energetics and by fitting the burst lightcurve with the cooling models of CM04 extended to low column depths, in 't Zand et al.~(2005) showed that this burst is consistent with pure helium ignition at $y\approx 10^{10}\ {\rm g\ cm^{-2}}$. They also showed that pure helium accretion at $\approx 0.01\ \dot M_{\rm Edd}$ gives ignition at $10^{10}\ {\rm g\ cm^{-2}}$ if most of the heat released in the crust flows outwards. Figure \ref{fig:basefluxhe} shows the ignition column depth for pure helium accretion as a function of base flux. At $0.01\dot m_{\rm Edd}$, $Q_b\approx 1\ {\rm MeV}$ per nucleon is required for ignition at $y\approx 10^{10}\ {\rm g\ cm^{-2}}$. Our ignition models in this paper show that this requires that the core neutrino emissivity not be enhanced over modified URCA. Either a slow cooling rate enhanced by a factor of 30, or fast cooling in the core give ignition column depths $> 10^{11}\ {\rm g\ cm^{-2}}$ at this accretion rate. An increase in accretion rate of factors of $>5$ over the assumed value is required to bring the enhanced cooling models into agreement. At low accretion rates, the ignition depth is most sensitive to core temperature, and is not very sensitive to crust properties. In particular, the crust neutrino emission plays a smaller role than for superbursts because of the lower crust temperatures.

Pure helium bursts are also observed at a similar accretion rate to superburst sources, from the ultracompact binary 4U~1820-30. In this case, we showed that the observed burst properties are again best fit by models which maximize the outwards flux from the crust. Models with Cooper pair emission from the crust give ignition depths, recurrence times and energies that are too large by a factor of 5. This is consistent with the previous conclusions of Bildsten (1995) and Cumming (2003), who did not consider the crust or core physics, but noted that the helium layer must be quite hot to achieve ignition at the depth inferred from observations. If the accretion rate is larger by a factor of $\gtrsim 2$ than inferred from the X-ray luminosity, these constraints are relaxed.  The uncertainty associated with the accretion rate can be bypassed by studying the burst energy as a function of recurrence time, as shown in Figure \ref{fig:et_he}, however, current constraints are limited.  Unlike at low accretion rates, the ignition conditions mainly depend on crust properties at high rates, giving a complementary view of the interior.

\subsection{Conclusions and Future Work}

The observational progress on long Type I X-ray bursts has opened up a new probe of accreting neutron star interiors, complementary to studies of isolated neutron stars (e.g.~Yakovlev \& Pethick 2004, Page et al.~2005) and accreting neutron stars in quiescence (Brown, Bildsten, \& Rutledge 1998; Colpi et al.~2001; Rutledge et al.~2002; Wijnands et al.~2002; Yakovlev et al.~2004). We find in this paper that the long Type I burst from 2S~0918-549, a pure helium accretor at $0.01\ \dot M_{\rm Edd}$, is best fit by models with core neutrino emissivity equivalent to modified URCA or smaller. For superbursts, and pure helium bursts from 4U~1820-30, which occur at higher accretion rates $\gtrsim 0.1\ \dot M_{\rm Edd}$, the ignition models limit both core and crust neutrino emission. In particular, neutrino cooling by Cooper pairing of neutrons in the crust leads to superburst ignition column depths that are too large. Either the Cooper pairing emissivity is much less than current calculations suggest\footnote{An alternative explanation is that these sources are not neutron stars but rather ``strange stars'' (e.g.~Alcock, Farhi, \& Olinto 1986). Strange stars do not have an inner crust which would naturally explain the lack of emission due to Cooper pairing of neutrons. This scenario is explored in Page \& Cumming 2005).} (see Appendix C for an assessment of the uncertainties in this process), or an additional heating source not included in current superburst ignition models is required in the accumulating fuel layer.

Our models can be improved in several respects. First, our ignition models use the one-zone criterion of Fujimoto, Hanawa, \& Miyaji (1981) and Bildsten (1998) to estimate the ignition column depth. Although this technique compares well with numerical simulations and normal mode analyses (e.g.~Woosley et al.~2003; Cooper \& Narayan 2005), time-dependent calculations of ignition should be carried out to calculate ignition conditions. These calculations are in progress (Halpin \& Cumming 2005, in preparation). Our cooling models for burst lightcurves assume instantaneous burning of the fuel, and cannot address the physics of the rise. Further studies of the superburst rise are needed. An additional motivation for this is to understand the observed precursors to superbursts, possibly due to triggering of an overlying H/He layer by the carbon runaway. 

The thermal models of the interior used here assume steady state accretion. In fact, superbursts have been observed from transient systems, in which case the core temperature will be lower than assumed in our models. Observations of quiescent cooling in KS~1731-260 (Rutledge et al.~2002; Wijnands et al.~2002), interpreted as cooling of the crust, imply  a cold core and high crust conductivity, exactly opposite to the conclusions from superburst ignition calculations, as emphasized by Brown (2004). The time-dependent calculations of Rutledge et al.~(2002) for this source indicate that the crust temperature reaches a maximum value of $\approx 2.5\times 10^8\ {\rm K}$, lower than the temperature in our steady-state models. Cooper pair cooling in the crust was not included in the models of Rutledge et al.~(2002), but is probably not important at these low temperatures. The lower crust temperatures may make achieving ignition at the inferred column depth for KS~1731-260 difficult, implying that an extra heating mechanism is required in superburst models. On the other hand, this may be related to the larger ignition column depth for KS~1731-260 compared to sources such as 4U~1636-54 for example. Further work on the time-dependent thermal profile in transiently accreting sources and the consequences for superburst ignition is needed.

\acknowledgements 
We thank L.~Bildsten, E.~Brown, and R.~Rutledge for helpful comments, and T.~Strohmayer and R.~Cornelisse for kindly supplying RXTE/PCA and BeppoSAX/WFC  lightcurves. We thank Achim Schwenk and Dima Yakovlev for discussions about the content of Appendix~C. AC acknowledges support from McGill University startup funds, an NSERC Discovery Grant, Le Fonds Qu\'eb\'ecois de la Recherche sur la Nature et les Technologies, and the Canadian Institute for Advanced Research. We thank S.~Woosley for making possible support for JM at the University of California, Santa Cruz through DOE grant No.\ DE-FC02-01ER41176 to the Supernova Science Center/UCSC. DP aknowledges partial support from a UNAM-DGAPA grant \#IN112502.


\begin{appendix}

\section{A. The Early Phase of the Superburst Lightcurve}

When fitting our time-dependent cooling models to observed superburst lightcurves, the inferred value of $E_{17}$ depends on the early part of the cooling curve, for times shorter than the thermal time of the fuel layer. Therefore it is important to understand the physics of this phase of the lightcurve. In this Appendix, we describe a simple steady-state model of the early cooling which highlights the physics, and gives us confidence in our numerical results.

CM04 showed that after the fuel burns, the early phase of the superburst lightcurve is set by a cooling wave which propagates inwards from the surface. At column depth $y$, there is a characteristic thermal timescale $t_{\rm therm}(y)\approx H^2/D$, where $H$ is the pressure scale height, and $D$ the thermal diffusivity. The thermal timescale grows with depth. After time $t$, the cooling wave has penetrated to a depth $y_b$ at which $t\approx t_{\rm therm}(y_b)$. At lower column depths $y<y_b$, the atmosphere has a constant heat flux with depth; at higher column depths $y>y_b$, the temperature profile has not evolved significantly from its initial state.

This picture suggests a simple model of the early phase of the lightcurve. We first make an analytic estimate for constant opacity in the layer, and then present numerical calculations integrating the true opacity profile. For constant opacity, the temperature profile in the constant flux region is given by integrating $F=(4acT^3/3\kappa)(dT/dy)$ from the surface. The radiative zero solution is
\begin{equation}\label{eq:A0}
T=4.0\times 10^9\ {\rm K}\ F_{24}^{1/4}y_{12}^{1/4}\left({\kappa\over 0.02\ {\rm cm^2\ g^{-1}}}\right)^{1/4},
\end{equation}
where we choose a typical value for the opacity. Equation (\ref{eq:A0}) gives the temperature profile in the constant flux region. However, at the base of the constant flux layer, the temperature remains close to its initial value, which is set by the nuclear energy release at the beginning of the flash. CB01 and CM04 calculated this to be $T_i=4.0\times 10^9\ {\rm K}\ y_{12}^{1/8}E_{17}^{1/2}(g_{14}/2.45)^{1/8}$. Combining these two conditions by using the radiative zero solution to set $T(y_b)=T_i$ gives the flux through the surface when the cooling wave has penetrated to depth $y_b$,
\begin{equation}\label{eq:A1}
F_{24}=y_{b,12}^{-1/2}E_{17}^2\left({\kappa\over 0.02\ {\rm cm^2\ g^{-1}}}\right)^{-1}\left({g_{14}\over 2.45}\right)^{1/2}.
\end{equation}
All that remains is to relate the depth of the constant flux region to the time after ignition. The thermal time is $t_{\rm therm}\approx H^2/D$, where $D=K/\rho c_P$ is the thermal diffusivity, giving
\begin{equation}
t_{\rm therm}={\rho c_P H^2\over K}={3c_P\kappa y^2\over 4acT^3},
\end{equation}
or taking $\kappa$ to be constant,
\begin{equation}\label{eq:A2}
t_{\rm therm}=11\ {\rm hours}\ {y_{12}^{3/2}\over E_{17}}\ \left({\kappa\over 0.02\ {\rm cm^2\ g^{-1}}}\right)\left({g_{14}\over 2.45}\right)^{-1/4}.
\end{equation}
Now setting $t=t_{\rm therm}$ and rewriting equation (\ref{eq:A1}) in terms of $t_{\rm therm}$ using equation (\ref{eq:A2}) specifies the flux evolution with time. We find
\begin{equation}
F_{24}\approx 2.2\ \left({t\over {\rm hrs}}\right)^{-1/3}E_{17}^{5/3}\left({\kappa\over 0.02\ {\rm cm^2\ g^{-1}}}\right)^{-2/3}\left({g_{14}\over 2.45}\right)^{5/12},
\end{equation}
which compares favorably to the empirical fit to the time-dependent simulations from CM04, $F_{24}\approx 2\ (t/{\rm hr})^{-0.2} E_{17}^{7/4}$.

\begin{figure}
\epsscale{0.5}
\plotone{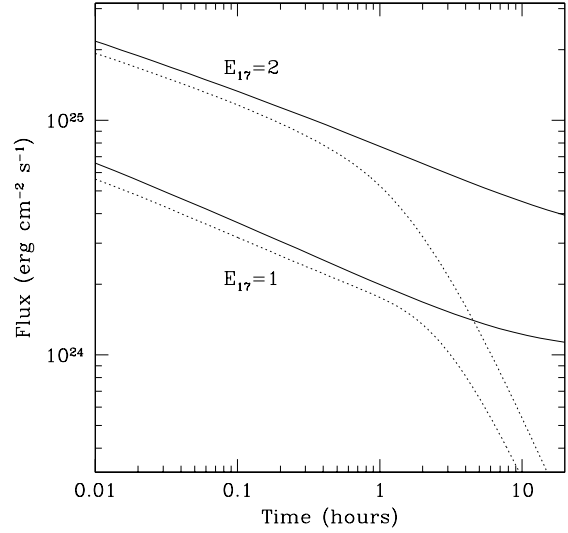}
\caption{Flux as a function of time for $E_{17}=1$ (lower curves) and $E_{17}=2$ (upper curves). The solid curves show the simple steady-state model described here; the dotted curves are the results of time-dependent simulations (with $y_{12}=1$).\label{fig:flux}}
\end{figure}

In fact, the opacity varies through the layer, typically by a factor of 5. We therefore repeat the same argument, but integrating numerically. For a given column depth $y_b$, we find the flux $F$ required so that a constant flux atmosphere has a base temperature which matches the initial temperature at that depth (set by the initial energy release). We then calculate the thermal time at column depth $y_b$, which tells us the time at which the cooling wave reaches $y_b$ or when this value of surface flux applies.  Figure \ref{fig:flux} shows the resulting lightcurves, compared with our time-dependent cooling models. The agreement is good, giving us confidence in the results of  our numerical models.

\begin{figure}
\epsscale{0.5}
\plotone{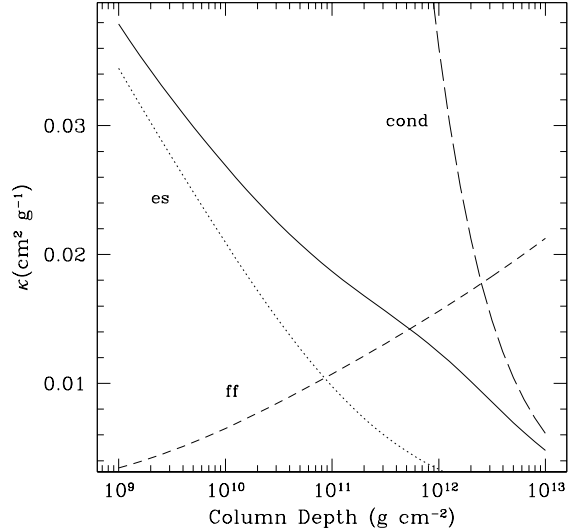}
\caption{Opacity (continuous line) as a function of depth immediately after burning for $E_{17}=2$. We also show the separate contributions to radiative opacity from electron scattering (es; dotted curve) and free-free absorption (ff; short-dashed curve), and the contribution from electron conduction (cond; long-dashed curve). At low densities, electron scattering dominates the opacity; at high densities, free-free absorption blocks radiative transport of heat, and electron conduction takes over.\label{fig:opac}}
\end{figure}

It is important to note that the radiative opacity can be important even close to the base of the burning layer\footnote{The plasma frequency $\omega_p$ in the burning layer is quite large, and one might therefore expect a suppression of radiative heat transport caused by the inability of photons with $\omega<\omega_p$ to propagate (we thank D.~Yakovlev for raising this issue). This is expected for temperatures $T\ll T_p$, where $T_p=\hbar\omega_p/k_B\approx 10^{10}\ {\rm K}\ (\rho_9Y_e)^{1/2}$. We have calculated the suppression factors for our models based on the modified Rossland mean opacity integrals given by Aharony \& Opher (1979) and van Horn (1992), and find that this effect is not important, changing the calculated cooling curves by of order 1\%. The smallness of the effect is firstly due to the fact that $T$ is always close to or greater than $T_p$ when radiative transport dominates electron conduction (at lower temperatures, $T_p/T$ is larger, but radiative heat transport is no longer important relative to conduction), and secondly because free-free absorption dominates the opacity in the region most likely to be affected (its $1/\omega^3$ frequency dependence leads to a preference for higher energy photons than electron scattering in the Rosseland mean integral, and so less suppression at a given temperature; Aharony \& Opher 1979).}, because of the high temperatures reached initially. Figure \ref{fig:opac} shows the opacity profile in the layer immediately after the fuel burns. The radiative opacity controls heat transport at lower densities, and has contributions from electron scattering and free-free absorption. At high density, the growing free-free opacity shuts off the radiative heat transport, and electron conduction takes over.  As the layer cools, the transition from radiative to conductive cooling happens at lower and lower densities, since $\kappa_{\rm cond}\propto T^2$, and $\kappa_{ff}\propto T^{-7/2}$, moving the intersection  point to lower density. A similar transition was noted in models of steady H/He burning by Schatz et al.~(1999), but at lower column depths because of the lower temperatures ($5$--$10\times 10^8\ {\rm K}$, in that case). For the free-free Gaunt factor, we use the fit of Schatz et al.~(1999) to the calculation of Itoh et al.~(1991). We have checked that the physical conditions in our models are within the regime of applicability of these calculations.

\section{B. Analytic models of the crust temperature profile}

Figure \ref{fig:all_he} shows that the flux entering the accreted layer (and therefore the burst ignition conditions) mainly depends on the core neutrino emissivity at low accretion rates, and the crust properties at high accretion rates. Brown (2000) computed the thermal structure of rapidly accreting neutron stars, and pointed out that the thermal structure of the crust could be understood using a simple analytic model. We take a similar approach here to understand the variation in the temperature profile as a function of accretion rate.

\subsection{Without neutrino cooling in the crust}

We first ignore neutrino emission from the crust, and assume that the heat released in the crust $Q_{\rm nuc}\approx 1.4\ {\rm MeV}$ per baryon is either radiated from the surface or conducted into the core and radiated as neutrinos. For accretion rates $\dot m\gtrsim 0.1\ \dot m_{\rm Edd}$, most of the energy enters the core. In this case, the peak temperature in the crust is whatever it needs to be  to conduct the heat into the core, given the thermal conductivity. If electron-ion collisions set the conductivity, as for a disordered crust, then
\begin{equation}
K=\left({\pi^2k_B^2Tn_e\over 3 m_\star}\right)\left({3\pi\hbar^3\over 4e^4m_\star Z}\right)
\end{equation}
(e.g.~Yakovlev \& Urpin 1980), where we assume a single species of ions with charge $Z$, $m_\star=E_F/c^2$ is the electron effective mass, and the second term is the inverse of the electron-ion collision frequency. We have set the Coulomb logarithm to unity, a reasonable approximation for our purposes.

We now solve $F=\rho K dT/dy$ in the inner crust to find the temperature profile. In the inner crust, the pressure is mostly from degenerate non-relativistic neutrons, giving $P\propto (Y_n\rho)^{5/3}$. Using this to integrate, we find
\begin{equation}
T_8\approx 16\ f_{in}^{1/2}\left({\dot m\over \dot m_{\rm Edd}}\right)^{1/2}\left({\rho_c\over 2\times 10^{14}\ {\rm g\ cm^{-3}}}\right)^{1/6},
\end{equation}
where we write the fraction of the energy released in the crust that flows inwards as $f_{in}$. To write this expression, we have assumed that the core temperature is much smaller than the maximum crust temperature, and that the density at the crust/core boundary is much greater than the density at the location of the temperature maximum (these approximations are good enough for our purposes). We have also taken $Y_n=0.8$, $Y_e=0.05$, and $Z=22$ in the inner crust. The accretion rate enters this formula because the inwards flux is $F=f_{in}Q_{\rm nuc}\dot m$.

Similarly, we can integrate into the crust from low densities. In the outer crust, degenerate relativistic electrons set the pressure. In this case, the temperature profile is given by
\begin{equation}\label{eq:outer}
T_8\approx 7.9\ \left({f_{\rm out}\over 0.1}\right)^{1/2}\left({Z^2/A\over 12}\right)^{1/2}\left({\dot m\over \dot m_{\rm Edd}}\right)^{1/2}\left[{\ln (P/P_0)\over 20}\right]^{1/2}
\end{equation}
(see also CB01), where $P_0$ is the outer pressure at which we start the integration, and we have assumed that the temperature there is small. 

Now matching the outer and inner temperatures, and assuming that $f_{\rm out}+f_{in}=1$ and $f_{\rm out}\ll 1$, we find
\begin{equation}\label{eq:appqb}
Q_b\approx Q_{\rm nuc}f_{\rm out}\approx 0.3\ {\rm MeV\ per\ nucleon}\ \left({Z^2/A\over 12}\right)^{-1}.
\end{equation}
The prefactor in this estimate is approximate. Brown (2000) gives a more careful analysis using more accurate power laws for the equation of state in the outer and inner crust. However, equation (\ref{eq:appqb}) nicely reproduces three properties of $Q_b$ that we find in the numerical calculations at $\dot m\gtrsim 0.1\ \dot m_{\rm Edd}$. First, Figures \ref{fig:qy} and \ref{fig:all_he} show that at these accretion rates, $Q_b$ is not very sensitive to $\dot m$. In equation (\ref{eq:appqb}), $\dot m$ drops out altogether because the temperature profiles in the inner and outer crust both scale in the same way with $\dot m$. Second, $Q_b$ is insensitive to the core neutrino emission as long as the maximum temperature in the crust is much greater than the core temperature. Thirdly, equation (\ref{eq:appqb}) shows that $Q_b$ is inversely proportional to $Z^2/A$. This was pointed out by Brown (2004) (the actual scaling is $Q_b\propto (Z^2/A)^{-0.6}$), and is the reason why the ignition conditions are much less sensitive to composition than found by CB01 (who varied $Z^2/A$ but kept $Q_b$ constant)\footnote{Note that an implicit assumption in this argument is that the composition of the inner crust does not change as $Z^2/A$ in the outer crust changes. In fact, this assumption is borne out by the calculations of Haensel \& Zdunik (2003).}.

At low accretion rates $\dot m<0.1\ \dot m_{\rm Edd}$, a large fraction of the heat released in the crust flows out rather than in, $f_{\rm out}\sim 1$, as shown in Figure \ref{fig:all_he}. The inner crust is almost isothermal in this case (see Fig.~[\ref{fig:prof_he}]), and the outwards flux is determined by the core temperature, which in turn is set by the core neutrino emission. This makes $Q_b$ more sensitive to the core neutrino emission, and less sensitive to crust properties, than at higher accretion rates. 

\subsection{With neutrino cooling in the crust}

We have ignored neutrino emission from the crust. However, as we have seen in the numerical calculations neutrino emission can have a dramatic effect on the temperature profile. One way to estimate when crust neutrino emission matters is to ask at what temperature the crust neutrino emission balances the energy release in the crust. For crust Bremsstrahlung, the neutrino emissivity is $Q_\nu\approx 0.3\ {\rm erg\ g^{-1}\ s^{-1}}\ T_8^6 L (1-Y_n) (Z^2/A)$, where $L$ is the Coulomb logarithm, and is insensitive to depth (see Fig.~\ref{fig:profq} and Haensel et al.~1996); we estimate a typical value $Q_\nu\approx 3\times 10^{12}\ {\rm erg\ cm^{-3}\ s^{-1}}\ T_8^6$, and a total crust luminosity  $L_\nu\approx 3\times 10^{30}\ {\rm erg\ s^{-1}}\ T_8^6$. The total nuclear energy release in the crust is $L_{\rm crust}\approx 10^{36}\ {\rm erg\ s^{-1}}\ (\dot m/\dot m_{\rm Edd})$. Therefore $L_\nu=L_{\rm crust}$ when 
\begin{equation}\label{eq:Tbrems}
T_8\approx 8\ (\dot m/\dot m_{\rm Edd})^{1/6},\hspace{2 cm}({\rm Bremsstrahlung})
\end{equation} 
or $T_8\approx 4$ for $0.01\ \dot m_{\rm Edd}$, and $T_8\approx 6.5$ for $0.3\ \dot m_{\rm Edd}$. This is in good agreement with Figures \ref{fig:prof_c} and \ref{fig:prof_he}, where neutrino emission plays an important role only for the hottest models in these figures. For Cooper pair neutrino emission in the crust, $Q_\nu\approx 7\times 10^{20}\ {\rm erg\ cm^{-3}\ s^{-1}}\ T_8^7\ (k_F/{\rm fm^{-1}})$ (Yakovlev et al.~1999). Using a height of $0.5\ {\rm km}$ for the emitting region gives $L_\nu\approx 3\times 10^{31}\ {\rm erg\ s^{-1}}\ T_8^7$. Again setting $L_\nu=L_{\rm crust}$, we find that neutrino emission will dominate the energy loss at the lower temperature
\begin{equation}\label{eq:cooperT}
T_8\approx 4.4\ (\dot m/\dot m_{\rm Edd})^{1/7}.\hspace{2 cm}({\rm Cooper\ pairs})
\end{equation}
This is in good agreement with Figure \ref{fig:prof_c3}. Ignition of carbon at column depths $<10^{12}\ {\rm g\ cm^{-2}}$ requires temperatures close to $6\times 10^8\ {\rm K}$ at the ignition point (see ignition curve in Fig.~\ref{fig:prof_c}). This is close to the maximum temperature for $\dot m=0.3\ {\rm \dot m_{\rm Edd}}$ without Cooper pairs, explaining why we need to go to higher accretion rates to get ignition at $y_{12}<1$. With Cooper pairs, the temperature at the ignition point is constrained to be $\approx 4\times 10^8\ {\rm K}$, giving ignition columns of $y_{12}\approx 10$.

\subsection{Limits on the outwards flux and ignition column depth}

The fact that the crust temperature is limited by neutrino emission from the crust gives an upper limit to the outwards flux at a given accretion rate. Substituting the limiting temperature from either equation (\ref{eq:Tbrems}) or (\ref{eq:cooperT}) into equation (\ref{eq:outer}) for the temperature profile of the outer crust, gives
\begin{equation}
Q_b\left({\dot m\over \dot m_{\rm Edd}}\right)<0.10\ {\rm MeV\ per\ nucleon}\ \left({\dot m\over \dot m_{\rm Edd}}\right)^{1/3}
\left({Z^2/A\over 12}\right)^{-1}\left[{\ln (P_1/P_0)\over 20}\right]^{-1}\hspace{2 cm}({\rm Bremsstrahlung})
\end{equation}
\begin{equation}
Q_b\left({\dot m\over \dot m_{\rm Edd}}\right)<0.043\ {\rm MeV\ per\ nucleon}\ \left({\dot m\over\dot m_{\rm Edd}}\right)^{2/7}
\left({Z^2/A\over 12}\right)^{-1}\left[{\ln (P_1/P_0)\over 20}\right]^{-1}\hspace{2 cm}({\rm Cooper\ pairs})
\end{equation}
where $P_1/P_0$ is the ratio of inner to outer pressures. These limits on the flux translate into limits on the  ignition column depth, by comparison with Figure \ref{fig:baseflux}. The limit on $Q_b$ from Cooper pairs severely constrains the carbon ignition depth. If $Q_b(\dot m/\dot m_{\rm Edd})<0.04$, then $y_{\rm ign}\gtrsim 5\times 10^{12}\ {\rm g\ cm^{-2}}$ for carbon. Cooper pair emission cools the crust so efficiently that its temperature is too low to achieve the required outwards heat flux for ignition at $y\approx 10^{12}\ {\rm g\ cm^{-2}}$ inferred from observations of superbursts.

\section{C. Neutron $^1$S$_0$ pairing and neutrino emission by the formation 
         of Cooper pairs
         \label{App:Tc}}

\subsection{Neutron $^1$S$_0$ pairing}

Given the importance of the neutrino energy losses due to the Cooper pair formation process found in this work, we describe in this Appendix in some detail the nature of the underlying physics and robustness of the ingredients needed for our models. First of all, the existence of a neutron drip regime is beyond doubt and results from the finite depth ($\sim 50$ MeV) and finite width ($\sim$ nucleus diameter) of the nuclear potential, i.e., this potential can only accomodate a finite, and small, number of bound states. With increasing density, the enormous Fermi energy of the electrons leads to  neutronization and the resulting large number of neutrons cannot be accomodated within the bound levels, i.e., neutrons {\em have to} drip at high enough density\footnote{The only exception to this can occur in a ``strange star'' made of self-bound strange quark matter, in which normal baryonic matter can exist only below the neutron drip density in the form of a thin outer crust  (Alcock, Farhi, \& Olinto 1986).}. These dripped neutrons form a degenerate Fermi liquid and, according to the Cooper theorem (Cooper 1956), they will unavoidably pair and become a superfluid if there is {\em any} attractive interaction between them (immediately afer the development of the BCS theory it was proposed that nucleons in nuclei must pair due to this mechanism; Bohr, Mottelson, \& Pine 1958). The important, and delicate, issue is the value of the pairing critical temperature $T_c$ which depends very sensitively on the strength of the pairing interaction. At the density range relevant for a neutron star crust the dominant attractive interaction between the dripped neutrons is in the spin-angular momentum channel $^1$S$_0$, which, fortunately is very well understood {\em in vacuum}.

\begin{figure}
\epsscale{0.5}
\plotone{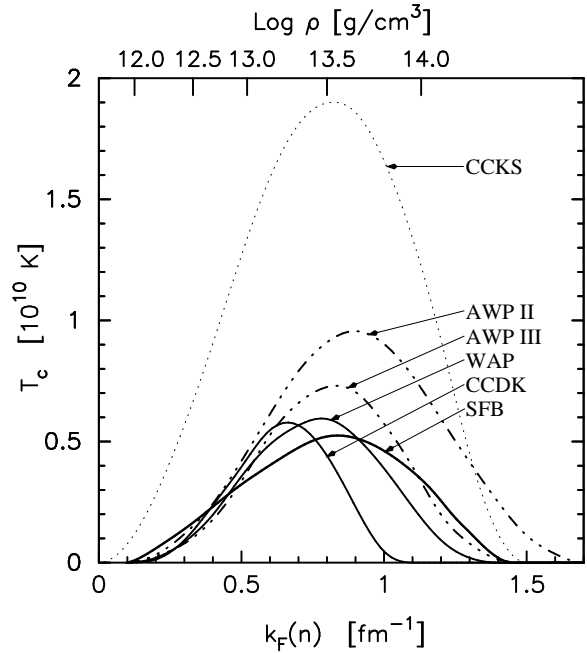}
\caption{Modern results for the neutron $^1$S$_0$ pairing critical temperature $T_c$
         as a function of the neutron Fermi momentum $k_F$, 
         from Table~\protect\ref{Tab:Tc-kf}.
         The curve labelled ``CCKS'' from an older calculation \protect (Chen et al.~1986), 
         in contradistinction to the other ones does not include medium polarization effects
         and is shown for illustration.
         The upper scale shows the corresponding densities for a crust chemical composition 
         as used in the present work}
\label{Fig:Tc-kf}
\end{figure}

In a medium, many-body effects alter the interaction and much effort has been dedicated to studying them (see, e.g., Lombardo \& Schulze 2001 for a review). At the densities corresponding to neutron star cores, there are still very large uncertainties in the size and density extent of the gaps, but in the low density regime of the crust the situation is much better. Many-body techniques used to calculate the size of the neutron $^1$S$_0$ gap have increased in sophistication with time and results obtained in the last fifteen years show a clear convergence. The most important effect turns out to be the polarization of the medium which, in some sense, screens the interaction between neutrons and results in a reduction of the gap. We plot in Figure~\ref{Fig:Tc-kf} results of the most reliable calculations (as well as one example of a calculation which explicitly did not include the effects of medium polarization for illustration of the importance of this effect). Numerical values are listed in Table~\ref{Tab:Tc-kf} as well as reference to the relevant works. Both the table and the figure illustrate the present uncertainties on $T_c$. A good indication of the convergence of  the models is that the most recent calculations of Schwenk, Friman, \& Brown (2003) incorporated significant new improvements and obtained values for $T_c$ very close to previous calculations. Given the range of temperatures in accreting neutron star crusts, $1 - 8 \times 10^8$ K, and considering that the Cooper pair neutrino process becomes negligible when $T < 0.2 \; T_c$ (see below), the important region is at neutron Fermi momenta smaller than 0.5 fm$^{-1}$, i.e., densities below $10^{13}$ g/cm$^3$ (Fig~\ref{Fig:Tc-kf}) where all $T_c$ curves we show are very close to each other. At higher densities the differences become significant but, fortunately, they do not seriously affect our results. For very low neutron densities one can use the following analytical solution (Schwenk 2004), which incorporates the medium polarization effect to calculate the zero temperature energy gap (see next paragraph),
\begin{equation}
\Delta(0) = (2/\mathrm{e})^{7/3} \; \epsilon_F \exp(\pi/2 k_F a_{nn})
\label{Eq:lowkf}
\end{equation}
where $a_{nn} = -18.5$ fm is the neutron-neutron scattering length and $\epsilon_F \equiv k_F^2/2 m_n$. The above formula is formally valid only when $|k_F a_{nn}| \ll 1$ but gives reasonable results even at $k_F \sim 0.1$ fm$^{-1}$. At all densities it is customary to calculate the critical temperature $T_c$ using the BCS result
\begin{equation}
\Delta(0) = 1.76 \; T_c \; .
\label{Eq:Tc}
\end{equation}
Even though this is not necessarily correct for calculations beyond the BCS approximation, it gives a credible value for $T_c$.

\begin{table}
\caption{Critical temperature $T_c$, vs neutron Fermi momentum $k_F$, 
         for five reliable calculations \label{Tab:Tc-kf}}
\begin{center}
\begin{tabular}{cccccc}
\hline
$k_F$ [fm$^{-1}$] &    \multicolumn{5}{c}{$T_c$ [10$^9$ K]}            \\
                  &  AWP II & AWP III &   WAP   &  CCDK   &  SFB       \\
\hline
\hline
       0.0        &   0.00  &   0.00  &   0.00  &   0.00  &   0.00     \\
$\,$   0.1$^\dagger$
                  &     -   &     -   &     -   &     -   &   0.19     \\
       0.2        &   0.33  &   0.26  &   0.20  &   0.13  &   0.59     \\ 
       0.3        &   1.18  &   0.79  &   0.86  &   0.92  &   1.38     \\ 
       0.4        &   2.44  &   1.78  &   1.98  &   2.37  &   2.31     \\ 
       0.5        &   4.20  &   3.36  &   3.56  &   3.96  &   3.23     \\ 
       0.6        &   6.13  &   5.27  &   4.88  &   5.47  &   4.02     \\ 
       0.7        &   7.91  &   6.59  &   5.67  &   5.67  &   4.75     \\ 
       0.8        &   9.10  &   7.25  &   5.93  &   4.42  &   5.21     \\ 
       0.9        &   9.56  &   7.05  &   5.21  &   2.31  &   5.14     \\ 
       1.0        &   9.03  &   5.74  &   3.89  &   0.46  &   4.62     \\ 
       1.1        &   7.71  &   3.96  &   2.31  &   0.00  &   3.76     \\ 
       1.2        &   5.93  &   1.98  &   0.92  &   0.00  &   2.55     \\ 
       1.3        &   4.15  &   0.79  &   0.20  &   0.00  &   1.25     \\ 
       1.4        &   2.50  &   0.13  &   0.00  &   0.00  &   0.20     \\ 
       1.5        &   1.12  &   0.00  &   0.00  &   0.00  &   0.00     \\ 
       1.6        &   0.36  &   0.00  &   0.00  &   0.00  &   0.00     \\ 
       1.7        &   0.00  &   0.00  &   0.00  &   0.00  &   0.00     \\
\hline
\end{tabular}
\end{center}
AWP II \& AWP III: Ainsworth, Wambach, \& Pines (1989),
WAP: Wambach, Ainsworth, \& Pines (1993),
CCDK: Chen et al.~(1993),
SFB: Schwenk, Friman, \& Brown (2003). \\
$^\dagger$ For values of $k_F$ this small, $T_c$ cannot be found from the above 
           references, except in the ``SFB'' case:           
           using Eq.~\protect\ref{Eq:lowkf} one obtains $T_c = 1.9\times 10^8$ K
           \protect (Schwenk 2004).
\end{table}

\begin{figure}
\epsscale{0.5}
\plotone{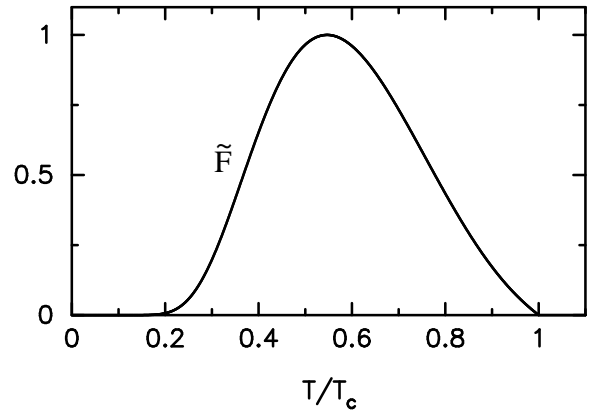}
\caption{Control function $\tilde{F}$ for Eq.~\protect\ref{Eq:PBF}}
\label{Fig:F-Delta}
\end{figure}

\subsection{Neutrino emission by Cooper pair formation}

The essence of pairing is the formation of a condensate of Cooper pairs, with the result that the energy of single-particle excitation changes from
\be
\epsilon(k) = \frac{\hbar^2 k^2}{2M_n} - \frac{\hbar^2 k_F^2}{2M_n}
= v_F \hbar (k-k_F)
\;\;\;\;\;  \mathrm{with} \;\;\;\;\;
v_F \equiv \frac{1}{\hbar}\left. \frac{d\epsilon}{dk}\right|_{k=k_F}
\ee
in absence of pairing to
\be
\epsilon_\mathrm{P}(k) = \sqrt{(v_F \hbar (k-k_F))^2 + \Delta(T)^2}.
\ee
in presence of pairing. 
The breaking of a Cooper pair therefore requires an energy of at least 
$2\Delta(T)$ while its formation generates that much energy.
At zero temperature there are no single-particle excitations, but at finite 
temperature $T<T_c$ the neutron fluid consists of two components, 
paired neutrons and excited ones, which are in thermal equilibrium and
there is constant formation and breaking of Cooper pairs. 
The energy liberated by the formation of a Cooper pair 
can be emitted in a neutrino-antineutrino pair 
(Flowers, Ruderman, \& Sutherland 1976; Voskresensky \& Senatorov 1987). 
For neutron $^1$S$_0$ pairing, the emissivity is
(Yakovlev, Kaminker, \& Levenfish 1999)
\be
Q_\nu^\mathrm{n \, ^1S_0} =
   1.5 \times 10^{22} \; \left[\frac{\hbar k_F}{M_n c}\right] \; 
    T_9^7 \; \tilde{F}(T/T_c) \;\; 
    \mathrm{erg \; cm^{-3} \; s^{-1}} \; .
\label{Eq:PBF}
\ee
The control function\footnote{Our normalized function $\tilde{F}$ is related 
to the $F$ of Yakovlev et al.~(1999) by $\tilde{F} = F/F_\mathrm{max}$  
with $F_\mathrm{max} = 4.313$.} 
$\tilde{F}$ is plotted in Figure \ref{Fig:F-Delta}. 
The shape of $\tilde{F}$ reflects the fact that when $T \sim T_c$ only a small
fraction of neutrons are in the paired state and the gap is small, 
resulting in a diminished emission of neutrinos with little energy, 
while when $T \ll T_c$ it becomes impossible to break pairs and hence pair 
formation does not occur anymore, resulting in a strong suppression of the 
emissivity approximately by a Boltzmann-like factor $\exp{-\Delta(0)/k_B T}$. 

This neutrino emission process is extremely efficient but not ``exotic'' in any 
way and has to be considered in any cooling neutron star model, even minimal 
(Page et al.~2004). Finally, we mention that a similar process occurs in the
laboratory when free electrons are injected into a superconductor, although 
the energy released is emitted predominantly in phonons (Schrieffer \& 
Ginsberg 1962) instead of neutrinos (!), leading to a pair formation rate 
in agreement with experiments (see, e.g., Carr et al.~2000 for recent results).

\end{appendix}


\begin{references}

\noindent
Aharony, U., \& Opher, R.\ 1979, \aap, 79, 27

\noindent
Ainsworth, T.~L., Wambach, J., \& Pines, D. 1989, Phys Lett B, 222, 173

\noindent
Alcock, C., Farhi, E., \& Olinto, A. 1986, \apj, 310, 261

\noindent
Augusteijn, T., van der Hooft, F., de Jong, J.~A., van Kerkwijk, M.~H., \& van Paradijs, J.\ 
1998, \aap, 332, 561 

\noindent
Baiko, D.~A., \& Yakovlev, D.~G. 1996, Astron.~Lett., 22, 708

\noindent
Bildsten, L. 1995, \apj, 438, 852

\noindent
Bildsten, L.\ 1998, NATO ASIC Proc.~515: The Many Faces of Neutron Stars., 419 

\noindent
Bildsten, L.\ 2000, American Institute of Physics Conference Series, 522, 359 

\noindent
Bohr, A., Mottelson, B. R., \& Pine, D. 1958, Phys. Rev., 110, 936

\noindent
Brown, E.~F. 2000, \apj, 531, 988

\noindent
Brown, E.~F. 2004, \apj, 614, L57

\noindent
Brown, E. F., \& Bildsten, L. 1998, \apj, 496, 915

\noindent
Brown, E.~F., Bildsten, L., \& Rutledge, R.~E. 1998, \apj, 504, L95

\noindent
Carr, G. l., Lobo, R. P. S. M., La Veigne, J., Reitze, D. H., \& Tanner, D. B. 2000, \prl, 85, 3001

\noindent
Caughlan, G.~R., \& Fowler, W.~A. 1988, At.~Data Nucl.~Data Tables, 40, 283

\noindent
Chen, J.~M.~C., Clark, J.~W., Dav\'e, R.~D., \& Khodel, V.~V. 1993, Nucl. Phys., A555, 59

\noindent
Chen, J.~M.~C., Clark, J.~W., Krotscheck, E., \& Smith, R.~A. 1986, Nucl. Phys. A 451, 509

\noindent
Colpi, M., Geppert, U., Page, D., \& Possenti, A. 2001, \apj, 548, L175

\noindent
Cooper, L.~N. 1956, Phys. Rev., 104, 1189

\noindent
Cooper, R.~L., \& Narayan, R. 2005, \apj, in press (astro-ph/0410462)

\noindent
Cornelisse, R., Heise, J., Kuulkers, E., Verbunt, F., \& in't Zand, J.~J.~M.\ 2000, \aap, 357, L21 

\noindent
Cornelisse, R., Kuulkers, E., in't Zand, J. J. M., Verbunt, F., \& Heise, J. 2002, \aap, 382, 174

\noindent
Cornelisse, R. et al. 2003, \aap, 405, 1033

\noindent
Cumming, A. 2003a, \apj, 595, 1077

\noindent
Cumming, A. 2003b, in "The Restless High-Energy Universe" (Amsterdam,
May 5-8, 2003), ed. E.P.J. van den Heuvel, J.J.M. in 't Zand, and R.A.M.J.
Wijers (astro-ph/0309626)

\noindent
Cumming, A., \& Bildsten, L. 2000, \apj, 544, 453

\noindent
Cumming, A., \& Bildsten, L. 2001, \apj, 559, L127 (CB01)

\noindent
Cumming, A., \& Macbeth, J. 2004, \apj, 603, L37 (CM04)

\noindent
Cumming, A., Morsink, S.~M., Bildsten, L., Friedman, J.~L., \& Holz, D.~E.\ 2002, \apj, 564, 343 

\noindent
Flowers, E., Ruderman, M., \& Sutherland, P. 1976, \apj, 205, 541

\noindent
Fushiki, I., \& Lamb, D.~Q.\ 1987a, \apj, 317, 368 

\noindent
Fushiki, I., \& Lamb, D. Q. 1987b, \apj, 323, L55

\noindent
Fujimoto, M.~Y., Hanawa, T., \& Miyaji, S.~1981, \apj, 247, 267

\noindent
Galloway, D.~K., Cumming, A., Kuulkers, E., Bildsten, L., Chakrabarty, D., \& Rothschild, 
R.~E.\ 2004, \apj, 601, 466 

\noindent
Gasques, L.~R., et al.~2005, \prc, in press (astro-ph/0506386)

\noindent
Gusakov, M.~E., Kaminker, A.~D., Yakovlev, D.~G., \& Gnedin, O.~Y.\ 2004, Astronomy Letters, 30, 759 

\noindent
Haensel, P., Kaminker, A.~D., \& Yakovlev, D.~G. 1996, \aap, 314, 328

\noindent
Haensel, P., \& Zdunik, J.~L. 1990, \aap, 227, 431

\noindent
Haensel, P., \& Zdunik, J.~L. 2003, \aap, 404, L33

\noindent 
in~'t Zand, J.~J.~M., Cornelisse, R., \& Cumming, A.\ 2004, \aap, 426, 257 

\noindent
in~'t Zand, J.~J.~M., Cumming, A., Verbunt, F., van der Sluys, M.~V., \& Pols, O.~R. 2005, \aap, in press (astro-ph/0506666)

\noindent
in~'t Zand, J.~J.~M., Kuulkers, E., Verbunt, F., Heise, J., \& Cornelisse, R.\ 2003, \aap, 411, L487 

\noindent
in~'t Zand, J.~J.~M., Verbunt, F., Kuulkers, E., Markwardt, C.~B., Bazzano, A., Cocchi, M., Cornelisse, R., Heise, J., Natalucci, L., \& Ubertini, P. 2002, \aap, 389, L43

\noindent
Itoh, N., Hayashi, H., Nishikawa, A., \& Kohyama, Y. 1996, \apjs, 102, 411

\noindent
Itoh, N., \& Kohyama, Y. 1993, \apj, 404, 268

\noindent
Itoh, N., Kuwashima, F., Ichihashi, K., \& Mutoh, H.\ 1991, \apj, 382, 636 

\noindent
Jones, P.~B. 2001, \mnras, 321, 167

\noindent
Jones, P.~B.\ 2004a, \mnras, 351, 956 

\noindent
Jones, P.~B.\ 2004b, \prl, 93, 221101 

\noindent
Juett, A.~M., Psaltis, D., \& Chakrabarty, D.\ 2001, \apjl, 560, L59 

\noindent
Kaptein, R.~G., in 't Zand, J.~J.~M., Kuulkers, E., Verbunt, F., Heise, J., \& Cornelisse, R. 2000, \aap, 358, L71

\noindent
Kitamura, H. 2000, \apj, 539, 888

\noindent
Kuulkers, E. 2002, \aap, 383, L5

\noindent
Kuulkers, E., et al.\ 2002, \aap, 382, 503 

\noindent
Kuulkers, E., 2003, in "The Restless High-Energy Universe" (Amsterdam,
May 5-8, 2003), ed. E.P.J. van den Heuvel, J.J.M. in 't Zand, and R.A.M.J.
Wijers (astro-ph/0310402)

\noindent
Kuulkers, E., in 't Zand, J.~J.~M., Homan, J., van Straaten, S., Altamirano, D., \& van der Klis, M. 2004, in ``X-Ray Timing 2003: Rossi and Beyond'', eds. P Kaaret, F.~K.~Lamb, \& J.~H.~Swank, American Institute of Physics Conf.~Proc. volume 714 (New York: Melville), p 253

\noindent
Lamb, D.~Q., \& Lamb, F.~K.\ 1978, \apj, 220, 291 

\noindent
Lewin, W.~H.~G., van Paradijs, J., \& Taam, R.~E.\ 1993, Space Science Reviews, 62, 223 

\noindent
Lewin, W. H. G., van Paradijs, J., \& Taam, R. E. 1995, in X-Ray Binaries,
ed. W. H. G. Lewin, J. van Paradijs, \& E. P. J. van den Heuvel (Cambridge:
CUP), 175

\noindent
Lombardo, U., \& Schulze, H.-J. 2001, in ``Physics of Neutron Star Interiors'', ed.~D.~Blaschke, N.~K.~Glendenning, and A.~Sedrakian, Lecture Notes in Physics, 578, 30 [astro-ph/0012209]

\noindent
Lorentz, C. P., Ravenhall, D. G., \& Pethick, C. J. 1993, Phys. Rev. Lett. 70, 379.

\noindent
Lyubarsky, Y., Eichler, D., \& Thompson, C. 2002, \apj, 580, L69

\noindent
Mackie, F. D., \& Baym, G. 1977, Nucl. Phys. A, 285, 332

\noindent
Muno, M.~P., Fox, D.~W., Morgan, E.~H., \& Bildsten, L.\ 2000, \apj, 542, 1016 

\noindent
Narayan, R., \& Heyl, J.~S.\ 2003, \apj, 599, 419 

\noindent
Nelemans, G., Jonker, P.~G., Marsh, T.~R., \& van der Klis, M.\ 2004, \mnras, 348, L7 

\noindent
Ogata, S., Ichimaru, S., \& van Horn, H.~M. 1993, \apj, 417, 265

\noindent
Page, D., \& Cumming, A. 2005, submitted to Ap. J. Lett. (astro-ph/0508444)

\noindent
Page, D., Geppert, U. \& Weber, F., submited to Nucl. Phys. A. (astro-ph/0508056)

\noindent
Page, D., Lattimer, J.~M., Prakash, M., \& Steiner, A.~A. 2004, \apjs, 155, 623

\noindent
Podsiadlowski, Ph., Rappaport, S., \& Pfahl, E.~D. 2002, \apj, 565, 1107

\noindent
Potekhin, A.~Y., \& Chabrier, G. 2000, \pre, 62, 8554

\noindent
Rutledge, R.~E., Bildsten, L., Brown, E.~F., Pavlov, G.~G., Zavlin, V.~E., \& Ushomirsky, 
G.\ 2002, \apj, 580, 413 

\noindent
Schatz, H., et al.\ 1998, \physrep, 294, 167 

\noindent
Schatz, H., et al. 2001, \prl, 86, 3471

\noindent
Schatz, H., Bildsten, L., \& Cumming, A. 2003a, \apj, 583, L87

\noindent
Schatz, H., Bildsten, L., Cumming, A., \& Wiescher, M. 1999, \apj, 524, 1014

\noindent
Schatz, H., Bildsten, L., Cumming, A., \& Ouellette, M. 2003b,
Nucl. Phys. A, 718, 247

\noindent
Schrieffer, J. R., \& Ginsberg, D. M. 1962, Phys. Rev. Lett., 8, 207

\noindent
Schwenk, A.~2004, talk at 12th International conference on Recent Progress in Many-Body Theory, Santa Fe [nucl-th/0411070]

\noindent
Schwenk, A., Friman, B., \& Brown, G.~E. 2003, Nuc Phys A, 713, 191

\noindent
Stella, L., Priedhorsky, W., \& White, N.~E. 1987, \apj, 312, L17

\noindent
Strohmayer, T. E., \& Bildsten, L. 2003, in Compact Stellar X-Ray
Sources, eds. W.H.G. Lewin and M. van der Klis (Cambridge: Cambridge
University Press) (astro-ph/0301544)

\noindent
Strohmayer, T. E., \& Brown, E. F., 2002, \apj, 566, 1045

\noindent
Strohmayer, T.~E., \& Markwardt, C.~B.\ 2002, \apj, 577, 337 

\noindent
Taam, R.~E., \& Picklum, R.~E.\ 1978, \apj, 224, 210 

\noindent
van Horn, H.~M.\ 1992, Bulletin of the American Astronomical Society, 24, 824

\noindent
Voskresensky, D.~N., \& Senatorov, A.~V. 1987, Sov. J. Nucl. Phys., 45, 411

\noindent
Wallace, R. K. \& Woosley, S. E. 1981, \apjs, 45, 389

\noindent
Wambach, J., Ainsworth, T.~L., \& Pines, D. 1993, Nucl. Phys., A555, 128

\noindent
Wijnands, R. 2001, \apj, 554, L59

\noindent
Wijnands, R., Guainazzi, M., van der Klis, M., \& M{\' e}ndez, M.\ 2002, \apjl, 573, L45 

\noindent
Woosley, S.~E., \& Taam, R.~E.~1976, \nat, 263, 101

\noindent
Woosley, S.~E., et al.\ 2004, \apjs, 151, 75 

\noindent
Yakovlev, D.~G., \& Haensel, P.~2003, \aap, 407, 259

\noindent
Yakovlev, D.~G., Levenfish, K.~P., Potekhin, A.~Y., Gnedin, O.~Y., \& Chabrier, G.\ 2004, 
\aap, 417, 169

\noindent
Yakovlev, D.~G., Kaminker, A.~D., \& Levenfish, K.~P.\ 1999, \aap, 343, 650 

\noindent
Yakovlev, D.~G., \& Pethick, C.~J. 2004, \araa, 42, 169

\noindent
Yakovlev, D.~G., \& Urpin, V.~A. 1980, Sov Astron, 24, 303



\end{references}
\end{document}